\documentclass[a4paper,12pt]{article}
\usepackage{amsmath}
\usepackage{amsfonts}
 \usepackage{graphicx} 
\usepackage[authoryear]{natbib}
\usepackage{xcolor}
\usepackage{float}
\usepackage{rotating} 
\usepackage[doublespacing]{setspace}
\usepackage[title]{appendix}
\usepackage[breaklinks=true]{hyperref} 
\usepackage{breakcites} 
\usepackage{url}
\usepackage{xr}
\usepackage{bm}
\usepackage{cleveref}
\RequirePackage{amsmath,amssymb}

\hypersetup{colorlinks=true,citecolor=blue, linkcolor=blue, urlcolor=blue}
\newtheorem{assumption}{Assumption}[section]
\newtheorem{theorem}{Theorem}[section]

\numberwithin{equation}{section}

\usepackage{threeparttable} 
\usepackage{physics} 
\usepackage{adjustbox} 
\hypersetup{colorlinks=true,citecolor=blue, linkcolor=blue, urlcolor=blue}
\usepackage{subcaption} 
\usepackage{ulem}
\usepackage{changes}
\usepackage{tcolorbox}

\newcommand{\Prob}{\ensuremath{\mathbb{P}}} 
\allowdisplaybreaks
\graphicspath{{Figures/}}
\newcommand{\blind}{1}

\begin{document}
	\def\spacingset#1{\renewcommand{\baselinestretch}%
		{#1}\small\normalsize} \spacingset{1}

\if1\blind
{
\title{\bf Sequential Cauchy Combination Test for Multiple Testing Problems with Financial Applications\thanks{%
 		We have received helpful comments and suggestions from Kris Boudt, Geert Dhaene, Frank Kleibergen, Nathan Lassance, Roberto Ren\`o, Olivier Scaillet, Rosnel Sessinou, Kristien Smedts, Steven Vanduffel, and the seminar and conference participants at 
 		KU Leuven, Université catholique de Louvain, 
 		the Netherlands Econometric Study Group (2022), the Quantitative Finance and Financial Econometrics Conference (2022), the Macquarie Financial Econometrics Workshop (2022), and the Computational and Financial Econometrics Conference 		(2022). Bouamara acknowledges support from the Flemish Research Foundation (FWO fellowship \#11F8419N) and the S\&B fund (Gustave Bo\"el -- Sofina fellowship).
 		Laurent acknowledges support from the French National Research Agency (reference: ANR-17-EURE-0020 and ANR-21-CE26-0007-01) and the excellence initiative of Aix-Marseille University - A*MIDEX.	Shi acknowledges research support from the Australian Research Council (project No. DE190100840).
 			\vspace{0.2cm}		\\ 
 		$^{\ddag}$ Nabil Bouamara, Louvain Institute of Data Analysis and Modeling in economics and statistics, Universit\'{e} catholique de Louvain;  	Email: nabil.bouamara@uclouvain.be.\\
 		$^{\ddag\ddag}$ S\'{e}bastien Laurent, Aix-Marseille University (Aix-Marseille School of Economics), CNRS \& EHESS, Aix-Marseille Graduate School of Management -- IAE; Email: sebastien.laurent@univ-amu.fr.\\ 
 		$^{\ddag\ddag\ddag}$ Shuping Shi, Department of Economics, Macquarie University; Email: shuping.shi@mq.edu.au. 
}
\vspace{0.1cm}
}

\author{Nabil Bouamara$^{\ddag}$, S\'{e}bastien Laurent$^{\ddag\ddag}$, Shuping Shi$%
^{\ddag\ddag\ddag}$ \\
}
 
\date{May 31, 2023} 

\maketitle
} \fi

\if0\blind
{
	\bigskip
	\bigskip
	\bigskip
	\begin{center}
		{\LARGE\bf Sequential Cauchy Combination Test for Multiple Testing Problems with Financial Applications}
	\end{center}
	\vspace{2cm}
} \fi

\vspace{-8mm}
\begin{spacing}{1.00} 
		\begin{abstract}
			We introduce a simple tool to control for false discoveries and identify individual signals 
			in scenarios involving many tests, dependent test statistics, and potentially sparse signals. 
			The	tool applies the Cauchy combination test recursively on a sequence of expanding subsets of $p$-values and is referred to as the sequential Cauchy combination test.  
			While the original Cauchy combination test aims to make a global statement about a set of null hypotheses by summing transformed $p$-values, our	sequential version 	determines which
			$p$-values trigger the rejection of the global null. 
			The sequential test achieves strong familywise error rate control, exhibits less conservatism compared to existing controlling procedures when dealing with dependent test statistics, and provides a power boost. 
			As illustrations, we revisit two well-known  
			large-scale multiple testing problems in finance for which the test statistics have either serial dependence or cross-sectional dependence, namely  monitoring drift bursts in asset prices and searching for assets with a nonzero alpha.  In both applications, the sequential 
				Cauchy combination test proves to be a preferable alternative. It overcomes many of the drawbacks  inherent to  inequality-based controlling procedures, extreme value approaches, resampling and screening methods,  and it improves  
				the power in simulations, leading to distinct empirical outcomes.
				\\
				
			
			 \vspace{-2mm}
			
		\end{abstract}
	
	\vspace{10mm}
	\end{spacing}

\newpage
\singlespacing 

\newpage


\section{Introduction}

There are many needle-in-a-haystack problems in empirical finance. There are tests to detect skilled funds \citep{barras2010false,giglio2021thousands}, nonzero alpha stocks  \citep{fan2015power}, explanatory factors  \citep{harvey2016and}, profitable technical trading rules \citep{bajgrowicz2012technical,sullivan1999data}, jumps and drift bursts in high-frequency asset prices \citep[][]{lee2007jumps, leemykland2012jumps,  bajgrowicz2015jumps,christensen2014fact,christensen2018drift},  to name but a few examples. 
These statistical tests have in common that, in order to detect a signal, they require applying the same  test repeatedly. This repetition arises either from the presence of a large cross-section of units (e.g., different funds, stocks, factors or trading rules) or because the test is required to be applied continuously over time (every minute or every day). 
It is commonly known that the simultaneous testing of multiple hypotheses is prone to a ``false discovery problem". As more and more tests are performed, an increasing number of them will be significant purely due to chance.\footnote{A well-known example of disregarding the multiple testing problem is so-called ``data snooping" or ``\textit{p}-hacking", which is a misuse of statistical testing: one  exhaustively searches for signals without compensating for the number of inferences being made 
	\citep[see e.g.,][and references therein]{giglio2021thousands}.} 
	To  counteract the inflation of false discoveries, researchers typically treat 
	the  hypotheses as a ``family" and set a threshold which controls a combined measure of error across all tests, rather than the Type I error of an individual test. 
	The procedures we focus on in this paper 
	address this issue by 
	imposing an upper bound, denoted as $\alpha$, on the probability of making at least one false discovery, thereby controlling the so-called ``familywise error rate"  \citep[see e.g.,][for reviews]{shaffer1995multiple,goeman2014multiple}.\footnote{Other procedures control, for example, the expected proportion of false discoveries or the so-called ``the false discovery rate" \citep[see e.g.,][]{benjamini1995,barras2010false,giglio2021thousands}.}

The dependence between test statistics plays a crucial role in the effectiveness of multiple testing corrections. 
In scenarios where the test statistics are independent and adhere to a standard normal distribution under the null hypothesis (as observed in certain jump tests, for instance), there exist well-established statistical solutions.\footnote{Popular approaches to address the false discovery problem in jump detection include 
	choosing a critical value corresponding to an extremely high quantile of the normal distribution 
	\citep[as in][]{andersen2007no} or 
	choosing another threshold based on the quantile of the  asymptotic distribution of the maximums of the test statistics \citep[as in][]{lee2007jumps,leemykland2012jumps}.} 
However, in many applications in economics and finance, assuming independence among the test statistics is implausible, because 
many popular test statistics are constructed from  overlapping rolling windows or are computed from stock returns that are likely to be driven by common factors. 
In scenarios with dependent test statistics, popular multiple testing corrections, such as those based the Gumbel distribution \citep[e.g.,][]{lee2007jumps} or statistical inequalities such as the Bonferroni correction and its subsequent improvements \citep{holm1979simple,hommel1988stagewise,hochberg1988sharper} protect against false discoveries, but are known to be overly conservative. The familywise error rate of these methods often turns out to be  much smaller than the desired upper bound $\alpha$. 
Simulation-based methods have also been used to
account for the observed correlation of the test statistics when setting a threshold  \citep[e.g.,][]{white2000,romano2005exact,romano2005stepwise}, but they are not ideal either. 
Aside from being computationally intensive,  they impose a strong parametric assumption on the dependence structure \citep[e.g., a Gaussian AR(1) process as in][]{christensen2018drift}, which could be misspecified. 

In this paper, we introduce a simple tool to control for false discoveries, while  being agnostic about the dependence among the test statistics. 
The solution we propose only uses raw $p$-values and
is built upon the Cauchy combination test of \citet{liu2020cauchy}, which tackles the issue of dependence in the test statistics from another perspective. 
Their global Cauchy combination (GCC) test  is grounded on a convenient theoretical property of Cauchy distributions, 
which states that  linear combinations of these variates behave similarly to a standard Cauchy variate at extreme tails, regardless of the dependence structure.  
Drawing from this insight, \citet{liu2020cauchy} propose a transformation of raw $p$-values, such that the transformed $p$-values follow a standard Cauchy distribution under the null hypothesis, and then construct a new test statistic as a linear combination of these transformed $p$-values, with its corresponding critical value derived from a Cauchy distribution. 
In doing so, they prove that the familywise error rate of the GCC test  converges to the nominal size $\alpha$ as the significance level $\alpha$ tends to zero, when all hypotheses are true and test statistics have arbitrary dependency structures. 
This test is well-suited to deal with the challenges posed by correlation,  high-dimensionality, and sparsity, but it is designed for  inferences about a global hypothesis. 
It is not obvious how statements about individual hypotheses are to be made with this procedure. 

We extend the pioneering work of \citet{liu2020cauchy} by introducing a sequential version of the Cauchy combination test to pinpoint the individual hypotheses that trigger the rejection of the global null, enabling the identification of  individual signals, such as, 
skilled  funds,  nonzero alpha stocks, explanatory factors, or timestamps of jumps and flash crashes.
We apply the GCC test recursively on expanding subsets of $p$-values, starting from the largest and progressing to the smallest $p$-value. This process generates a sequence of Cauchy combination test statistics. The $p$-values associated with these test statistics are computed based on a standard Cauchy distribution. Individual violations are detected when the corresponding $p$-value is lower than a predefined threshold $\alpha$. 
We refer to this new testing procedure as the sequential Cauchy combination (SCC) test, which inherits all the convenient theoretical properties of the GCC test, including being agnostic about the dependence structure. 
We prove that the SCC test achieves strong familywise error rate control as the significance level $\alpha$ tends to zero, regardless of whether the number of individual hypotheses is fixed or infinite. Moreover, 
compared to the benchmark procedures, the familywise error rate of the SCC test is closer to the theoretical upper bound, which boosts the power and helps to better identify the individual  signals. 
 
To showcase the advantages of the sequential Cauchy combination test, we revisit two 
multiple testing problems in financial econometrics that exhibit  non-trivial correlation structures in the test statistics, high dimensions, and sparse signals, which are common challenges in the field.

\begin{itemize}
\item In the first example, we revisit the drift burst hypothesis of \cite{christensen2018drift}, which aims to identify explosive trends in  stock prices. 
The drift burst test statistic relies on ultrahigh-frequency data and is applied multiple times within a trading day. 
The test statistics are constructed from overlapping rolling  windows and exhibit serial dependence. Drift bursts are rare events, and the strength of this signal varies over time.  

\item In the second example, we test for multiple nonzero alphas within the \citet{fama2015five} five-factor model framework. If the model fully explains asset returns, the estimated ``alphas" should be statistically indistinguishable from zero.
The presence of unknown common factors generates strong cross-sectional  dependencies among the test statistics \citep[see e.g.,][]{giglio2021thousands}. 
Nonzero alphas are typically rare and weak \citep[see e.g.,][and references therein]{fan2015power}. 
\end{itemize}

To assess 
the robustness of the SCC test against different   forms of dependence,  
we conduct two sets of simulation studies. 
The first set involves directly generating  test statistics with different correlation matrices. 
The second set involves generating data from a specific underlying process and computing a sequence of test statistics from the data, mimicking the situation in real-world empirical applications. 
Specifically, we simulate log prices of financial assets using a continuous-time drift bursting process in Example 1, and excess returns from a factor model in Example 2. 
The main findings from these simulations highlight that the SSC test outperforms other  multiple testing corrections, including statistical inequality-based approaches, methods based on extreme value theory, resampling, and screening approaches.  Despite its simplicity, the SCC test demonstrates superior properties 
in terms of  minimizing conservativeness and maximizing successful detections. 

The rest of the paper is organized as follows.
 Section \ref{secPrelims} introduces the general notation, definitions, and introduces both the global and sequential Cauchy combination tests. 
 Section \ref{secSims} illustrates the finite sample
 performance of the sequential Cauchy combination test in a simulation experiment with different types of correlations, relative to other multiple testing corrections.
 Sections \ref{secApplDriftBurst} and \ref{secApplFan}  revisit the two financial applications. 
 Section \ref{secConc} concludes. 
Appendix \ref{sec:proof} contains the proofs. 
The Online Supplement provides detailed information on the benchmark procedures, along with additional descriptions and simulations of the drift burst test and the nonzero alpha test.

\section{Multiple Hypothesis Testing with Correlated Test Statistics} \label{secPrelims}

In this section, we first introduce the notation and terminology used throughout the article regarding multiple hypothesis testing. 
We then review the global Cauchy combination test of \citet{liu2020cauchy}, followed by our sequential version of the Cauchy combination test.  

\subsection{Setting}

Let $H_{i}$ denote the $i^{\text{th}}$ null hypothesis of interest, with $i=1,...,d$. Here, $d$ denotes the total number of individual hypotheses, and $\mathcal{H}_{0}$ denotes the collection of null hypotheses of interest. 
To test the $d$ hypotheses, we can use the associated vector of test statistics $\bm{X}=(X_{1},X_{2},\ldots,X_{d})^{^{\prime }}$, one for each hypothesis being tested, or the corresponding raw $p$-values $p_{1},\ldots ,p_{d}$. The test statistics can be independent or dependent. 
For many popular tests, such as those described in Section \ref{secApplDriftBurst}, the test statistics are constructed from rolling windows and exhibit strong serial correlation.

To ensure the validity of individual hypothesis testing, it is common practice to control the probability of falsely rejecting a single hypothesis that is true (known as a false positive or Type I error) at a pre-specified nominal $\alpha$-level. 
However, when dealing with a large number of hypotheses, the issue of multiplicity arises: 
if the Type I error of each individual test is controlled at the $\alpha$-level, 
the probability of having at least one false positive conclusion rises well above $\alpha$. 

A classical global test circumvents the issue of multiplicity by replacing multiple tests with a single test. The corresponding global null hypothesis, denoted as $\mathcal{H}_0 = \bigcap_{i=1}^{d} H_{i}$, assumes that all elementary hypotheses are true, and the alternative hypothesis posits that at least one  hypothesis is false. 
In the context of monitoring specific events like jumps or drift bursts within a fixed time period, such as a day, the global null hypothesis would reflect the absence of any such event occurring within that given timeframe. Although global tests serve their purpose by aggregating effects, they may not provide the means to differentiate among individual hypotheses. In the field of financial econometrics, we are often interested in precisely timestamping drift bursts or identifying skilled fund managers, which requires a more granular analysis beyond the scope of global tests. 

Let $\mathcal{T}$ denote the set of true hypotheses,  $\mathcal{F}$  denote the set of false hypotheses,  and $\mathcal{R}$ denote the set of rejected hypotheses. 
The set of true and false hypotheses are unknown. 
A statistical test selects hypotheses to reject based on empirical data, and the corresponding set of discoveries in $\mathcal{R}$ 
should coincide with $\mathcal{F}$ as much as possible, while controlling the probability of making false discoveries. 
The objective of many multiple testing corrections is to control the familywise error rate (FWER), which constrains the probability of at least one false rejection within a family, denoted as $P[\mathcal{T} \cap \mathcal{R} \neq \varnothing]$. 
Ideally, the multiple testing correction should ensure that the FWER is not greater than the upper bound $\alpha$, while striving to keep it as close to $\alpha$ as possible. 
We concentrate on strong control of the FWER, allowing for the presence of some false hypotheses ($\mathcal{F} \neq \varnothing$), rather than weak FWER control, which assumes that all hypotheses of interest are true (i.e., $\mathcal{T} = \mathcal{H}_0$).

\subsection{Global Cauchy combination test}
\label{sec:CC}

The global Cauchy combination (GCC) test examines the global null hypothesis. 
The GCC test statistic is constructed from raw  $p$-values of the individual test statistics $X_i$, which are uniformely distributed between $0$ and $1$ under the  null hypothesis. 
The core idea of this test is first to transform these uniformly distributed $p$-values into standard Cauchy variates using the transformation formula $\tan \{(0.5-p_{i})\pi \}$, and then construct a new test statistic by taking the  weighted sum of these transformed $p$-values. 
The new test statistic is denoted by $\tilde{T}$ and is defined as: 
\begin{equation}
	\label{eqCauchyStatistic}
	{\normalsize \tilde{T}=\sum_{i=1}^{d}w_{i}\tan \{(0.5-p_{i})\pi \},} 
\end{equation}
in which the $w_{i}$'s are non-negative weights that sum up to one. Throughout the paper, the weights $w_{i}$ are set to $1/d$, for $i=1,\ldots,d$, following \citet{liu2020cauchy}.

When the raw $p$-values are independent or perfectly dependent, the new test statistic \eqref{eqCauchyStatistic} has a standard Cauchy distribution under the null, because the family of Cauchy densities is closed under convolutions. Even in cases of general dependence (whether weak, mild, or strong), the correlation structure has minimal impact on the tail behavior of the test statistics due to the heavy tails of the Cauchy distribution. 
Specifically, \citet{liu2020cauchy} prove that: 
\begin{equation}
	\lim_{h\rightarrow \infty }\frac{\Pr\left( \tilde{T}>h\right) }{\Pr\left(
		C>h\right) }=1,  \label{eq:tail}
\end{equation}
in which $C$ is a standard Cauchy random variable, subject to certain regularity conditions on the test statistic.

The result expressed in  \eqref{eq:tail} suggests that, under the global null hypothesis, the tail of the Cauchy combination test statistic is approximately Cauchy distributed, under arbitrary dependence structures, so that a $p$-value of the Cauchy combination test, denoted 
$\widetilde{p}$, can  be calculated from the standard Cauchy distribution. Suppose that we observe $\tilde{T}=t_{0}$, then: 
\begin{equation}
	\label{eqCauchyPval}
	\widetilde{p}=\frac{1}{2}-\frac{\arctan t_{0}}{\pi }. 
\end{equation}

Using the GCC $p$-values \eqref{eqCauchyPval}, the tail result in \eqref{eq:tail} can be equivalently stated as the actual size converging to the nominal size $\alpha$ as the significance level tends to zero:  
\begin{equation}
	\lim_{\alpha\rightarrow 0 }\frac{\Pr\left( \widetilde{p} \leq \alpha\right) }{\alpha}=1. \label{eq:wFWER}
\end{equation} 
 The approximation is particularly accurate for small $\alpha$'s, which are of particular interest in large-scale testing problems such as Examples 1 and 2 in sections \ref{secApplDriftBurst} and \ref{secApplFan}.
Importantly, the GCC test achieves weak familywise error rate control  regardless of the underlying correlation structure.

\begin{figure}[!h]
	\caption{The minimal impact of dependence on the tails of the GCC test statistic}
	\label{figCauchyPvalsAR}
	\centering
	
	\par
	
	\subfloat[${\theta} = 0.2$ ]{{\includegraphics[width=.40\textwidth,angle =
			-90,scale=0.70]{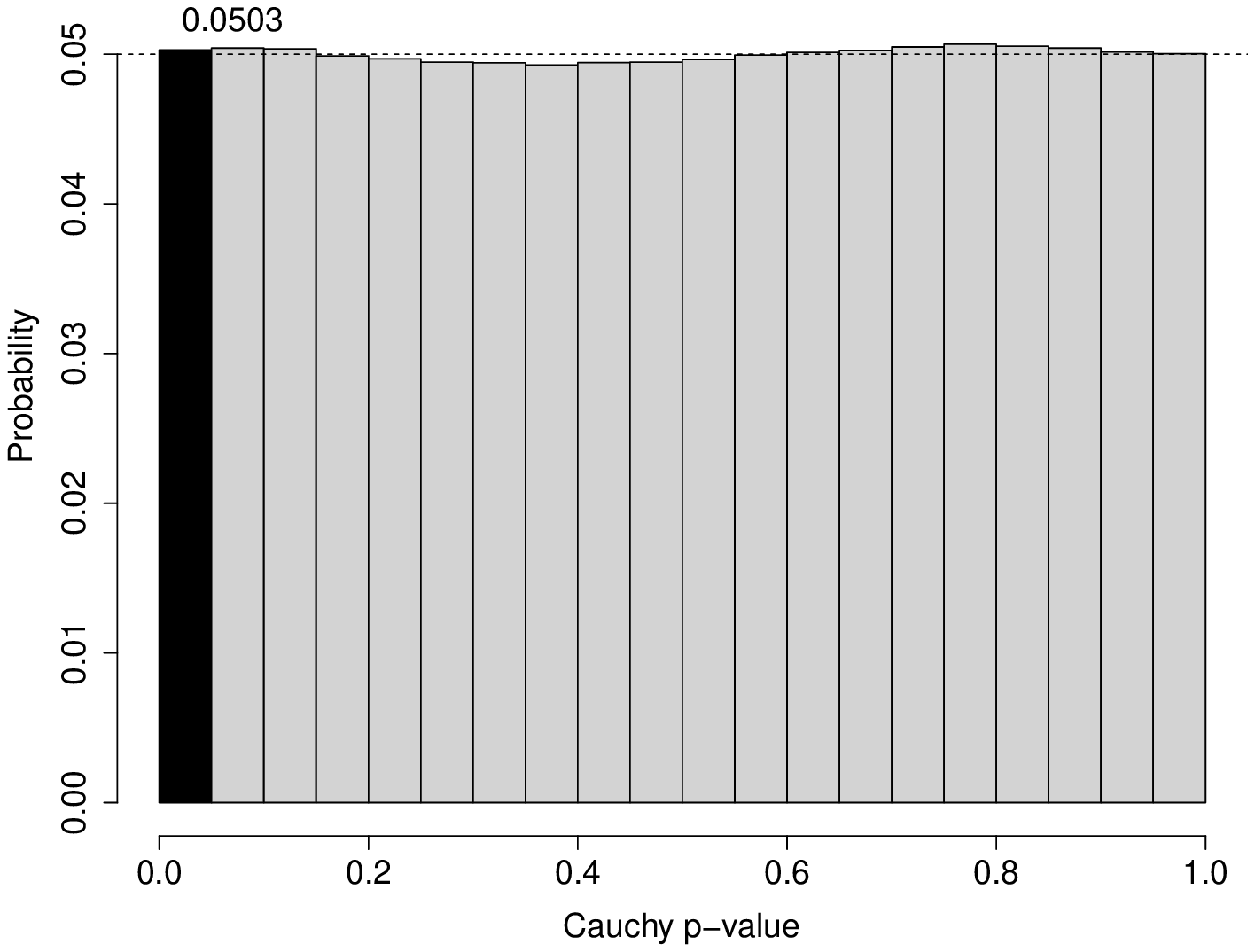} }} 
	\subfloat[$\theta = 0.4$ ]{{\includegraphics[width=.40\textwidth,angle =
			-90,scale=0.70]{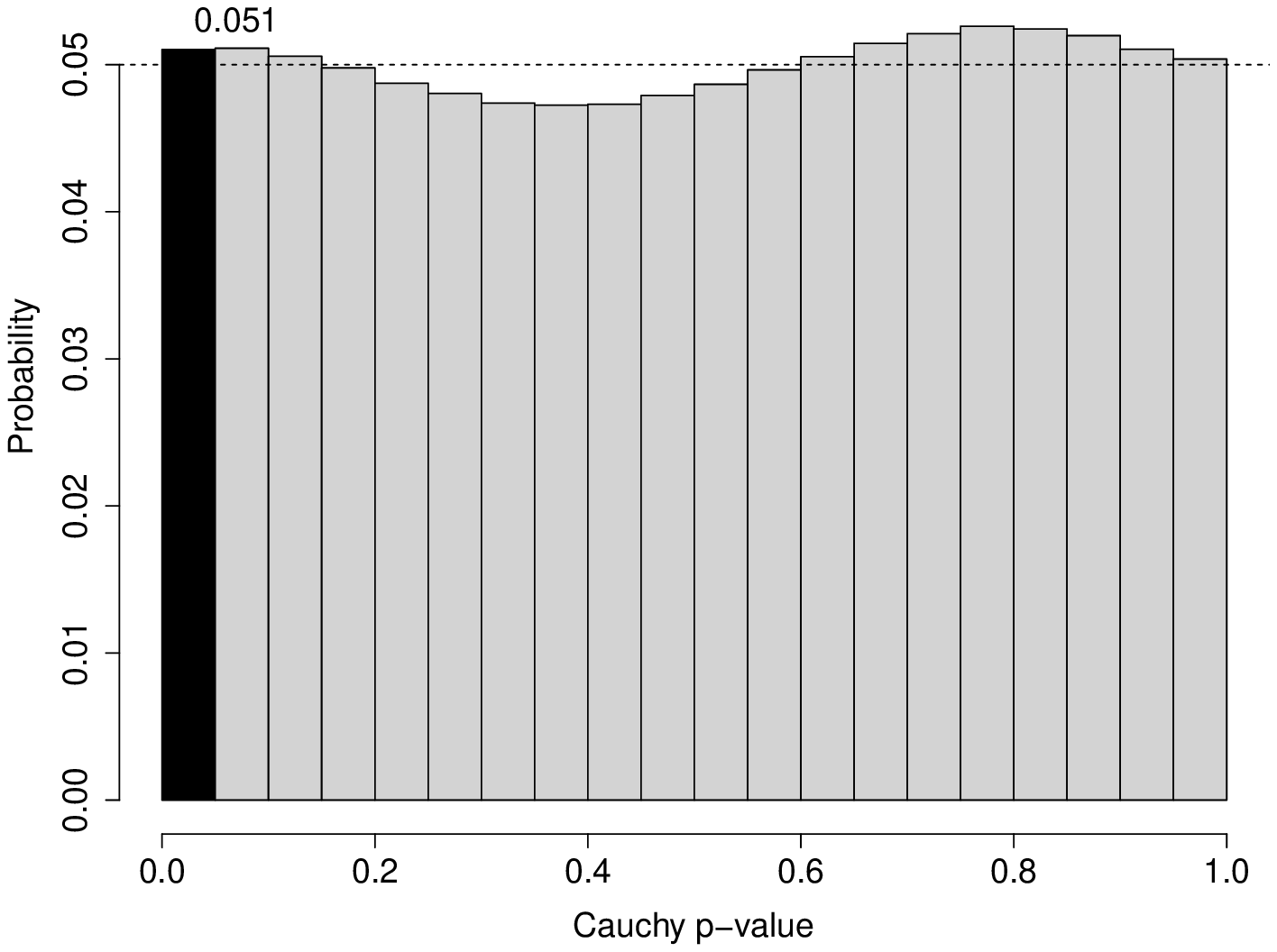} }} 
	
	\vspace{0.4cm}
	
	\subfloat[$\theta = 0.8$ ]{{\includegraphics[width=.40\textwidth,angle =
			-90,scale=0.70]{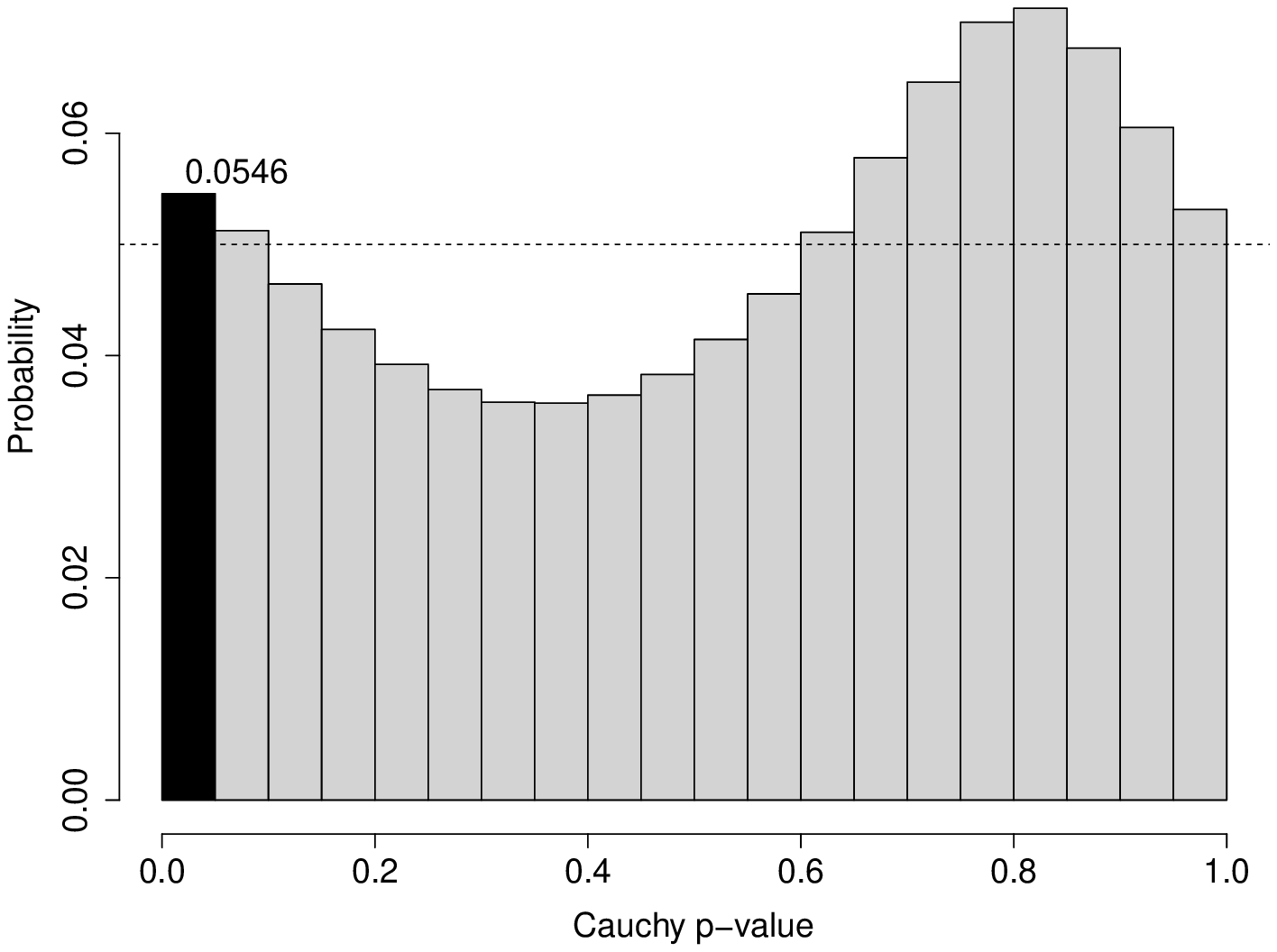} }} 
	\subfloat[$\theta = 0.95$]{{\includegraphics[width=.40\textwidth,angle =
			-90,scale=0.70]{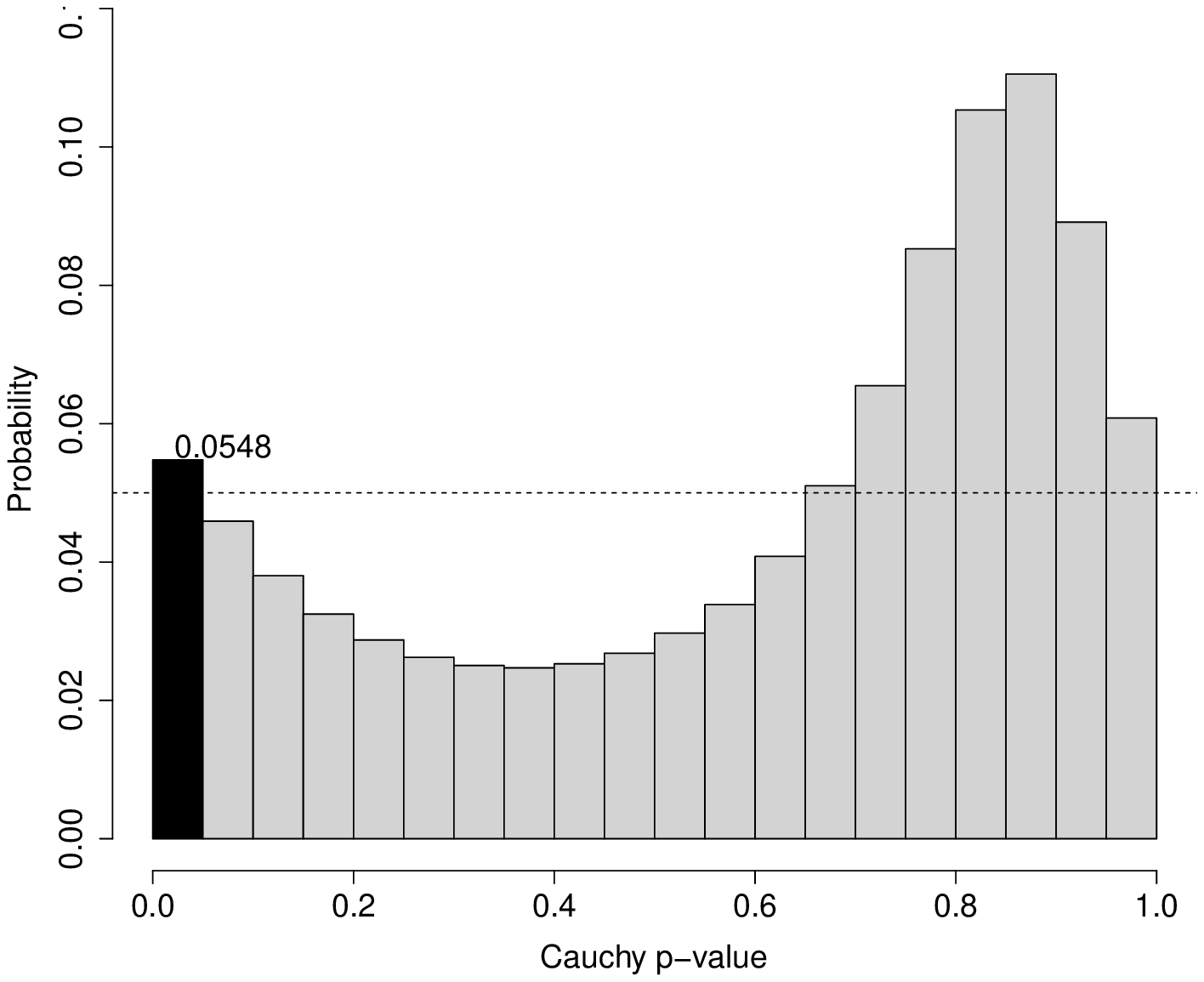} }} 
	
	\par
	\begin{minipage}{1.0\linewidth}
		\begin{tablenotes}
			\small
			\item {
				\medskip
				Note: We plot histograms of GCC $p$-values \eqref{eqCauchyPval} for various correlation strengths. The individual test statistics are drawn from a $d$-variate normal distribution $N_d(\bm{0}, \bm{\Sigma})$ with $\bm{\Sigma}= (\sigma_{ij})$ and $d=300$. The diagonal elements of the covariance matrix $\sigma_{ii}=1$ for all $i=1,\ldots,d$ and the off-diagonal elements  $\sigma_{ij} = \theta^{\vert i-j \vert}$ for $i\neq j$, with $\theta = 0.2, 0.4,0.8,0.95$. 
				The simulation is repeated $10^7$ times. The simulated GCC $p$-values are sorted into bins 
				with a width of  0.05. 
				We highlight the first bin in black and 
				add a text note with the probability of $p$-values being in the first bin. }
		\end{tablenotes}
	\end{minipage}
\end{figure}

Figure \ref{figCauchyPvalsAR} illustrates that while the dependence among individual test statistics may affect the null distribution of the GCC test statistic \eqref{eqCauchyStatistic}, its influence on the tail is minimal. 
To illustrate this point, we simulate a vector of  test statistics  $\bm{X}$ from a $d$-variate normal distribution with correlation matrix $\bm{\Sigma}$, \textit{i.e.}, $N_d(\bm{0}, \bm{\Sigma})$ with $\bm{\Sigma} = (\sigma_{ij})$ and $d = 300$. 
The diagonal elements $\sigma_{ii}=1$ for all $i=1,\ldots,d$ and the off-diagonal elements $\sigma_{ij} = \theta^{\abs{i-j}}$ for $i \neq j$, where $\theta$ takes the values of $0.2, 0.4, 0.8, 0.95$. 
We repeat the simulation $10^7$ times, and calculate two-sided $p$-values, the GCC test \eqref{eqCauchyStatistic} and the GCC $p$-value \eqref{eqCauchyPval} for each draw. 
The  histogram of the $10^7$ GCC $p$-values is plotted in Figure \ref{figCauchyPvalsAR}. 
When the autocorrelation is low (i.e., $\theta=0.2$), the distribution of $p$-values resembles a uniform distribution. As the autocorrelation increases, a pothole in the middle and a bump at the end of the histogram appear. However, regardless of magnitude of the autoregressive parameter, the percentage of $p$-values falling into the first bin remains consistently around $5$\%, as guaranteed by the limit result described in \eqref{eq:wFWER}. 

\subsection{Sequential Cauchy Combination Test}

The main contribution of this paper is the introduction of the sequential Cauchy combination test, which extends the GCC test of \citet{liu2020cauchy} to make statements on the elementary hypotheses. 
To facilitate this, the raw $p$-values are sorted in ascending order, denoting them as $p_{(1)}\leq p_{(2)}\leq \ldots \leq p_{(d)}$, where  $H_{(1)},H_{(2)},\ldots,H_{(d)}$ correspond to their respective null hypotheses.
For the purpose of testing hypothesis $H_{\left( i\right) }$, we calculate a Cauchy combination test statistic, denoted as  ${\normalsize \tilde{T}}_{\left( i\right) }$, using a subset of $p$-values running from $p_{(i)}$ to $p_{(d)}$ as:  
\begin{equation}
{\normalsize \ \tilde{T}%
		_{\left( i\right) }=\sum_{j=i}^{d}w_{j}\tan \{(0.5-p_{(j)})\pi \}
	} \text{ \ with \ } w_j =  \frac{1}{d-i+1},
	\label{eq:CC_mt}
\end{equation}
where $w_j$ represents the weight assigned to each $p$-value in the subset.
The corresponding $p$-value is computed as: 
\begin{equation}\label{eq:SCCp}
\widetilde{p}_{(i)}=\frac{1}{2}-\frac{\arctan \tilde{T}_{\left(i\right) }}{\pi }.
\end{equation} 
We reject the $i$th null hypothesis $H_{(i)}$ if  $\widetilde{p}_{(i)}\leq\alpha$. 
Similar to the step-up procedure introduced by \citet{hommel1988stagewise}, the SCC test leverages power across hypotheses: the test statistic $\tilde{T}_{(i)}$ is computed using the raw $p$-values associated with $\mathcal{H}_0^{(i)}=\bigcap_{j=i}^{d} H_{(j)}$. 

A more prescriptive description of the SCC testing procedure is as follows: 

\medskip

\begin{tcolorbox}
\begin{minipage}{1\linewidth}
	\textbf{SCC algorithm}	
	
	\begin{enumerate}
	
	\item Calculate raw $p$-values $p_1, p_2,\ldots, p_d$ corresponding to the null hypotheses $H_{1}, H_{2},\ldots, H_{d} $.%
	
	\item Order the raw $p$-values in ascending order, 	$p_{(1)},p_{(2)},\ldots,p_{(d)}$, with their  corresponding  ordered null hypotheses $H_{(1)},H_{(2)},\ldots,H_{(d)}$.
	
	\item Calculate the SCC test statistic $\tilde T_{(i)}$ and the transformed Cauchy $p$-values $\widetilde{p}_{(i)}$ from a subset of the ordered $p$-values $\left\{p_{(j)}\right\} _{j=i}^{d}$ using \eqref{eq:CC_mt} and \eqref{eq:SCCp}, respectively, for $i=1,\ldots,d$.
	
	\item Construct the rejection set $\mathcal{R}=\left\{H_{\left(i\right)} : \widetilde{p}_{(i)}\leq \alpha\right\}$. 
\end{enumerate}
\end{minipage}
\end{tcolorbox}

\begin{figure}[!h]
	\caption{Mechanics of the sequential Cauchy Combination test}
	\label{figSequentialCauchyIllustration}\centering	
	\par

	\subfloat[Raw
	$p$-values]{{\includegraphics[width=.31\textwidth,angle = -90]{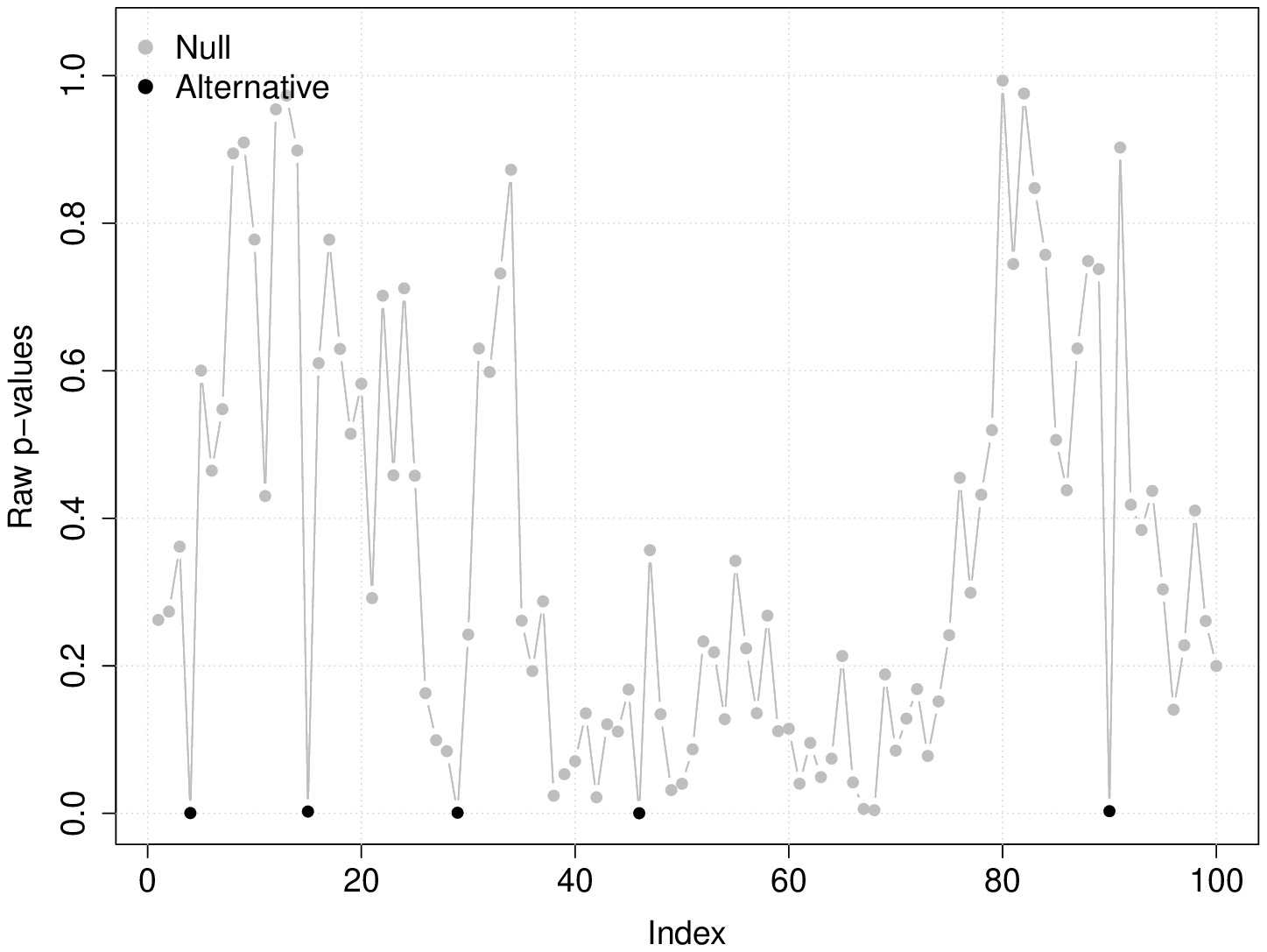} }}
	\subfloat[Ordered raw $p$-values]{{\includegraphics[width=.31\textwidth,angle =
			-90]{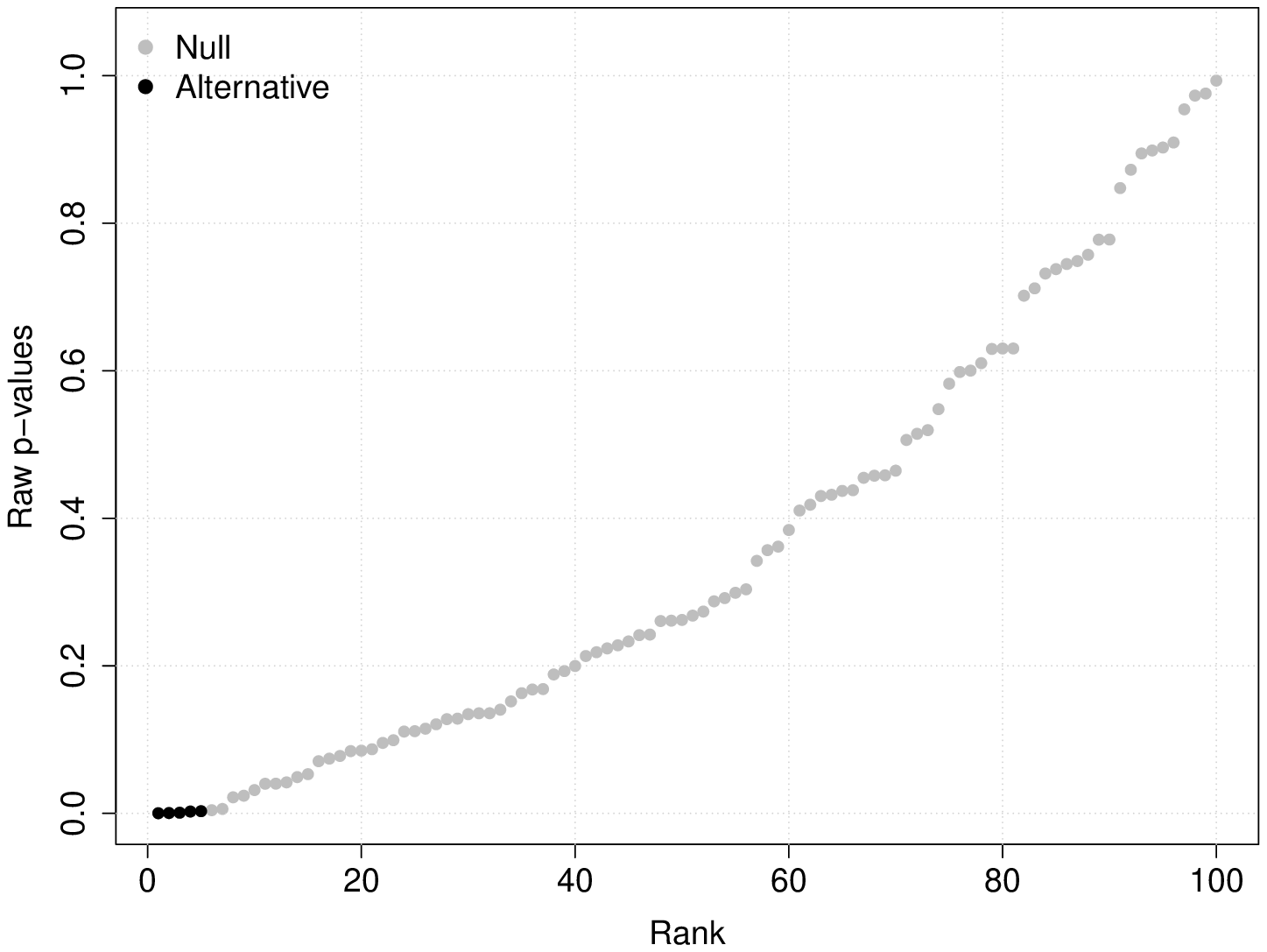} }} 
	
	\vspace{0.4cm}	
	
	\subfloat[SCC test statistics
	]{{\includegraphics[width=.31\textwidth,angle = -90]{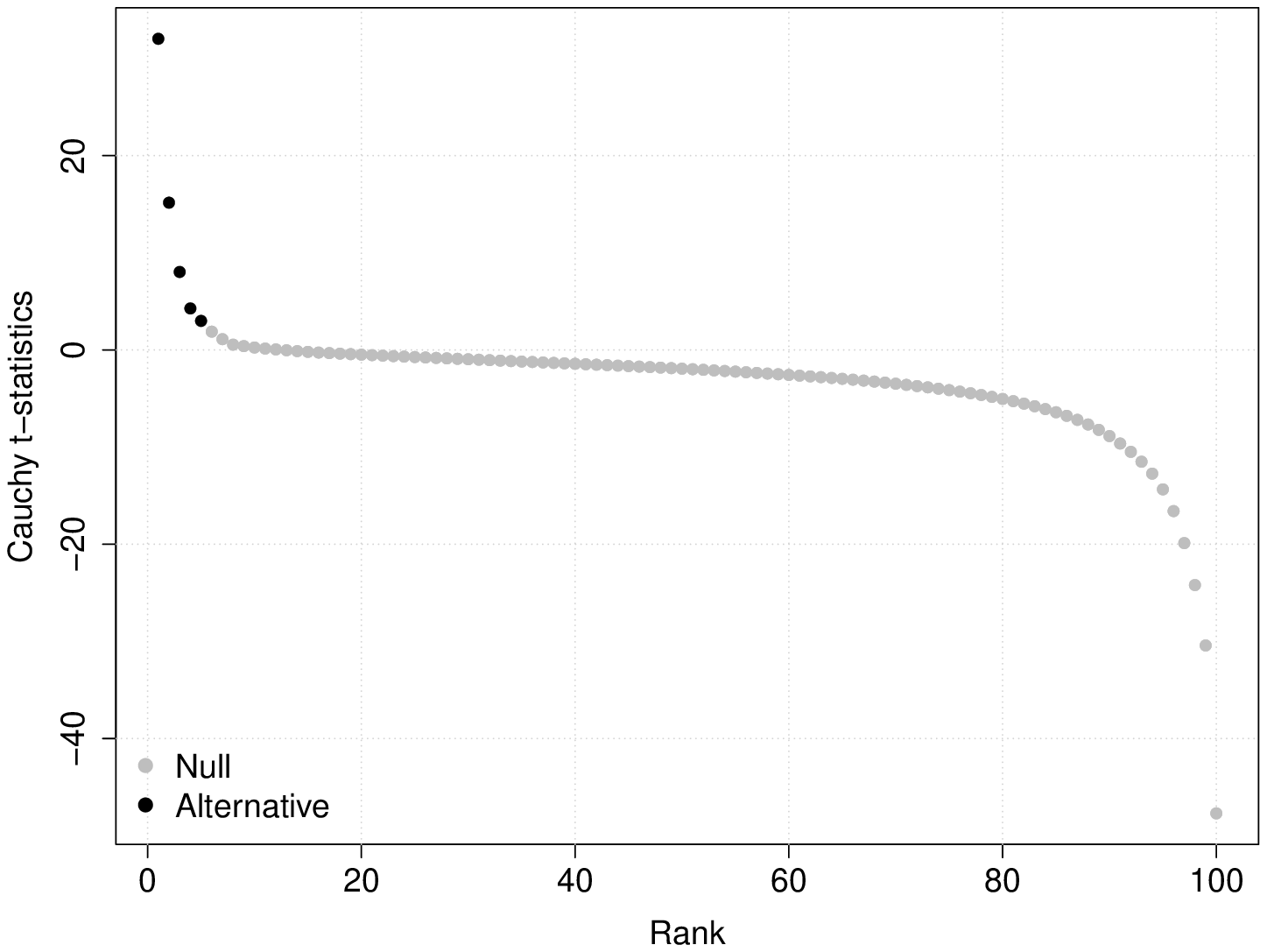}}}
	\subfloat[SCC 
	$p$-values]{{\includegraphics[width=.31\textwidth,angle = -90]{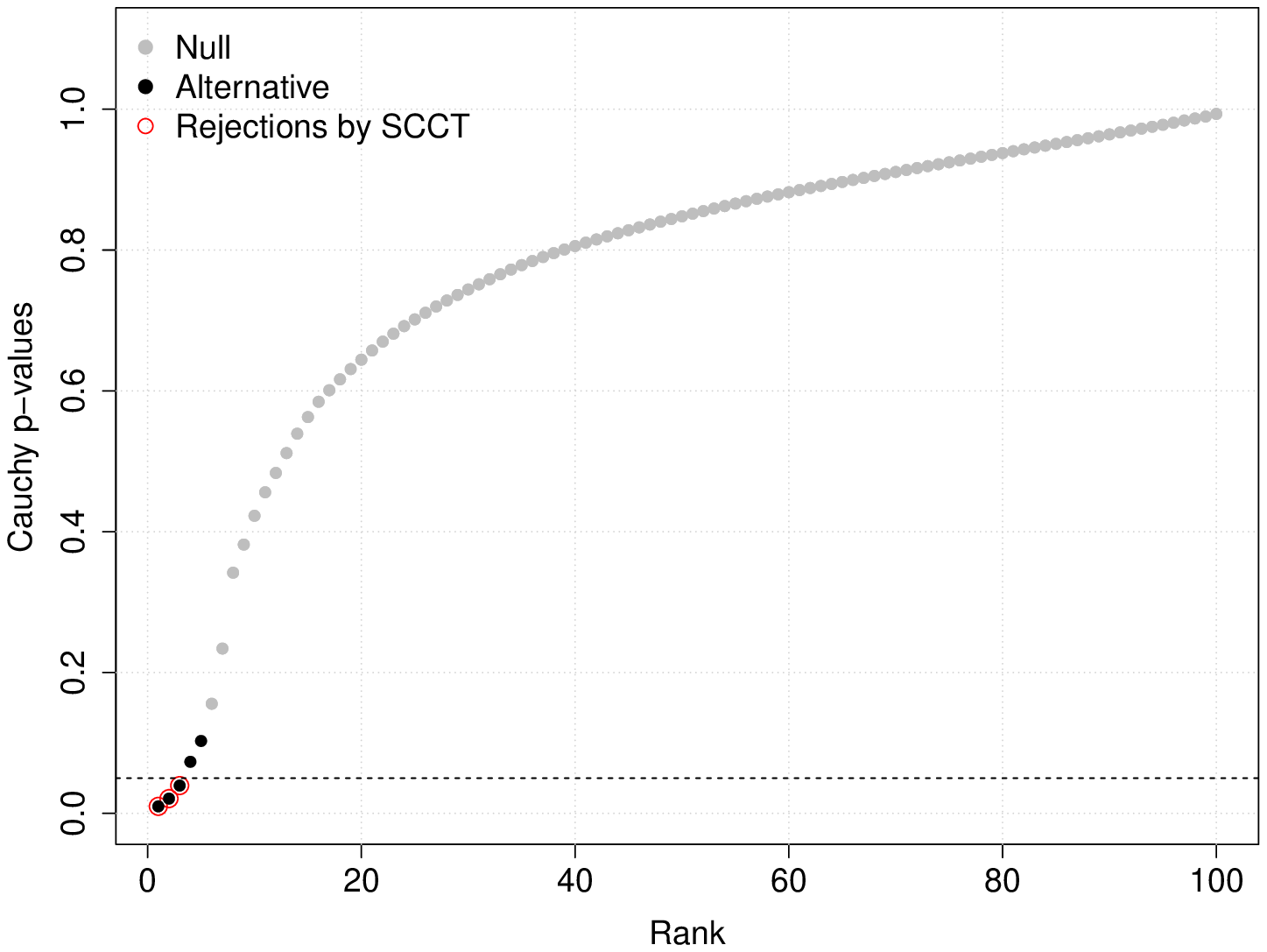}}}

	\begin{minipage}{1.0\linewidth}
		\begin{tablenotes}
			\small
			\item {
				\medskip
				Note: 
				The test statistics are simulated from $N_{d}(\bm{0},\bm{\Sigma})$ as in Figure \ref{figCauchyPvalsAR}, with $d=100$, $\theta=0.9$ and $5\%$ signals. The strength of the signal is $\pm2.806$, with its sign identical to that of the test statistic under the null.  
				The horizon line in panel (d) is the 5\% significance level.
			}
		\end{tablenotes}
	\end{minipage}
\end{figure}

Figure \ref{figSequentialCauchyIllustration} illustrates the mechanics of sequential Cauchy combination procedure using a simulated sequence of test statistics. 
The top row of the figure shows the raw and ordered 
$p$-values. Most of the observations correspond to the null hypothesis (represented by grey dots), while a few observations correspond to the alternative hypothesis (represented by black dots). 
The data-generating process is the same as the one used in Figure \ref{figCauchyPvalsAR}, where $\theta=0.9$ and $d=100$. 
We add constant signals (nonzero mean) for five out of the 100 hypotheses,  with a signal strength of $\pm2.806$. 
The sign of the signal aligns with the sign of the test statistic under the null, such that the signal always amplifies the magnitude of the test statistic. 
The bottom row of the figure plots the sequential Cauchy combination test statistics \eqref{eq:CC_mt} and their corresponding $p$-values \eqref{eq:SCCp}. In particular,  the bottom right panel shows that the SCC $p$-values $\widetilde{p}_{(i)}$ decrease as $i$ decreases from $d$ to $1$. In this example, the SCC test rejects three out of the five alternative hypotheses and does not reject any under the null hypothesis. These rejections correspond to the 4$^\text{th}$, 29$^\text{th}$ and  46$^\text{th}$ hypotheses in the top left panel. 
It is worth noting that the smallest SCC $p$-value corresponds to the $p$-value of the GCC test in \eqref{eqCauchyStatistic}, which performs the test on the largest set of hypotheses. 

The sequential Cauchy combination testing procedure requires two assumptions. 
Let $\bm{X}=(X_{1},X_{2},\ldots,X_{d})^{^{\prime }}$ represent the vector of test statistics. 

\begin{assumption}
	\label{ass1} (1) The original test statistics $(X_{i},X_{j})$, for any $1\leq
	i<j\leq d$, follow a bivariate normal distribution; (2) $E\left( \bm{X}%
	\right) =0$. 
\end{assumption}
The requirement of bivariate normality in Assumption \ref{ass1} is a condition weaker than joint normality, enabling the procedure to be applicable for high-dimensional settings. 
When the dimension $d$ increases at a certain rate with the sample size, the test statistics $\bm{X}$ may not jointly converge to a multivariate normal distribution due to a slower rate of convergence \citep[see][and references therein]{liu2020cauchy}, and assuming joint normality becomes unrealistic in such settings. 

 \citet[][]{liu2020cauchy} show through simulations that the global Cauchy approximation remains accurate even when the normality assumption is violated, and follows a multivariate Student-$t$ distribution (with four degrees-of-freedom) instead. 
For a showcase example in finance where the test statistics are Student-$t$ distributed, we refer the reader to Section \ref{secApplFan}. 

\begin{assumption}
	\label{ass2} Let $\mathbf{\Sigma }=corr\left( \bm{X}\right) $. 
	(1) The	largest eigenvalue of the correlation matrix $\lambda _{\max}\left( \mathbf{%
		\Sigma }\right) \leq C_0$ for some constant $C_0>0$; 
	(2) $\max_{1\leq i<j\leq
		d}\left\{ \sigma _{i,j}^{2}\right\} \leq \sigma _{\max }^{2}<1$ for some
	constant $0<\sigma _{\max }^{2}<1$, where $\sigma _{i,j}$ is the $\left(
	i,j\right) $ element of $\mathbf{\Sigma }$.
\end{assumption}
Assumption \ref{ass2} on the correlation matrix becomes relevant when the number of hypotheses $d$ diverges to infinity. 
It imposes two conditions: boundedness of the largest eigenvalue of the correlation matrix and the absence of perfectly correlated test statistics. 
The conditions are frequently encountered in high-dimensional settings and general enough to encompass a wide range of tests.\footnote{However, this assumption excludes very strong dependence and latent factors shared by the test statistics in high-dimensional settings. Investigating the relaxation of this assumption is left for future research.}

\begin{theorem}\label{thm}
	Under Assumption \ref{ass1} for a fixed $d$  and Assumptions \ref{ass1} and \ref{ass2} for $d=o(h^{\eta})$ with $0<\eta<1/2$, as $\alpha\rightarrow 0$,  the probability of the SCC testing procedure making at least one false rejection converges to $\alpha$, i.e., 
\begin{equation}\label{eq:conv}
\lim_{\alpha\rightarrow 0}\Pr\left\{\mathcal{R} \cap \mathcal{T}=\emptyset\right\} \rightarrow \alpha.
\end{equation}
\end{theorem}
The proof of Theorem \ref{thm} is provided in Appendix \ref{sec:proof}.  The theoretical result in \eqref{eq:conv} for the SCC testing procedure stands in stark contrast to statistical inequality-based controlling procedures covered in the Online Supplement, which have the property: 
$$\Pr \left\{ \mathcal{R}\cap \mathcal{T\neq \emptyset }\right\} \leq \alpha.$$  See \cite{goeman2010sequential} for a discussion of their theoretical properties. These inequality-based controlling procedures ensure that the likelihood of making at least one false discovery is bounded above by the pre-specified significance level, $\alpha$.  Consequently, the SCC procedure exhibits less conservatism compared to inequality-based controlling procedures. 

\section{Simulations}
\label{secSims}

In this section, we compare the performance of the SCC test against several popular multiple testing corrections in a simulation study, considering different forms of dependence. 
The benchmark procedures include four inequality-based approaches: the Bonferroni correction and its subsequent improvements proposed by  \citealp{holm1979simple}, \citealp{hommel1988stagewise} and \citealp{hochberg1988sharper}, as well as the Gumbel approach.  
Detailed discussions of these benchmark procedures can be found in the Online Supplement.  

\subsection{Under the null hypothesis}
\label{ssecAccuracy}

We assess the statistical performance of the different multiple testing corrections under the null hypothesis. 
To measure the empirical familywise error rate, we conduct $S=10^4$ replications for each method $m$, and calculate  $\widehat{FWER}_m$ as follows: 
\[
\widehat{FWER}_m=\frac{1}{S}\sum_{s=1}^{S} \bm{1}\left( \min_{i\in\left\{1,2,\cdots,d\right\}} \left\{p_{(m,i)}^{s}\right\}\leq\alpha\right),
\]
where $\bm{1}(.)$ is the indicator function, and $p_{(m,i)}^{(s)}$ represents the $p$-value of the $i$th hypothesis for method $m$ in the $s$th replication. 
When the test statistics exhibit strong dependence, we expect the  $\widehat{FWER}$ of the SCC test to be closer to the nominal level $\alpha$ compared to the other procedures. 

Under the null hypothesis, the test statistics $\bm{X}$ are generated from a $d$-variate normal distribution with zero mean and covariance matrix $\bm{\Sigma}$, i.e., $N_d(\bm{0}, \bm{\Sigma})$. 
We set the dimension $d$ to 100. 
The diagonal elements of the covariance matrix, $\sigma_{ii}$, are all equal to $1$, for $i=1,\ldots,d$. 
The off-diagonal element,  $\sigma_{ij}$ with $i\neq j$, adhere to three specific  specifications. 
\begin{itemize}
	
	\item Model 1. Exponential decay: $\sigma_{ij} = 	\theta^{\abs{i-j}}$ with $\theta= 0.2, 0.4, 0.6,$ $0.8, 0.90, 0.95$.
	
	\item Model 2. 	Polynomial decay: $\sigma_{ij} = \frac{1}{0.7 + \abs{i - j}^\theta}$ with $\theta =	1.0, 1.5, 2.0, 2.5$. 
			
	\item Model 3.	Block-diagonal:  $\bm{\Sigma} = \text{diag}\{A_1,\ldots, A_{d/10}\}$, for which each diagonal block $A_k$ is a $10 \times 10$ equi-correlation matrix with its off-diagonals $\sigma_{ij} = \theta$ and $\theta = 0.1, 0.3, 0.5, 0.7, 0.9$. 
	
\end{itemize}

Models 1 and 2 also appear in \citet{liu2020cauchy}. 
The exponentially decaying correlation structures in Model 1 are frequently observed in time series and financial econometrics. 
For instance, in Section \ref{ssecDB}, the sequence of drift burst test statistics constructed from overlapping rolling windows, exhibits an autoregressive process  and has an exponential decaying covariance structure, as shown by \citet{christensen2018drift}. 
The block-diagonal structure in Model 3 is commonly used 
when testing high-dimensional factor pricing models  \citep[see e.g., the Monte Carlo experiments in][]{fan2015power} and emulates a cross-sectional dependence structure.

Table \ref{tabSizeGlobalSequentialOtherMethods} shows the superior performance of the SCC test in the presence of correlated test statistics under the null hypothesis. 
The empirical familywise error rate of the SCC test, reported in the last column, remains close to the nominal level $\alpha = 5\%$ across various correlation structures, demonstrating its robustness. These findings align with the theoretical discussions presented in Section \ref{secPrelims}. 
In contrast, the inequality-based procedures and the Gumbel method exhibit greater conservatism, as evidenced by their substantially lower FWERs (although slight variations exist depending on the correlation pattern). 
The SCC test stands out as the only  controlling procedure with an empirical FWER close to the nominal level (5\%) across all three  types of correlation structures in the test statistics.

\begin{table}[htb!] 
	\centering
	\caption{Empirical FWERs (in\%) of the controling procedures}
	\label{tabSizeGlobalSequentialOtherMethods}
	\begin{adjustbox}{max width=\textwidth}
		\begin{tabular}{c ccccccc}
			\hline
			\multicolumn{1}{c}{$\theta$}
			&\multicolumn{1}{c}{Bonferroni}
			&\multicolumn{1}{c}{Holm} 
			&\multicolumn{1}{c}{Hommel}  
			&\multicolumn{1}{c}{Hochberg} 
			&\multicolumn{1}{c}{Gumbel}
			&\multicolumn{1}{c}{SCC} 
			\\
			\hline
			\multicolumn{7}{c}{\textbf{Model {1}:} Exponential decay} \\	
			0.2  & 4.68 & 4.68 & 4.68 & 4.68 & $\underline{3.30}$ &  4.94 \\ 
			0.4 & 4.88 & 4.88 & 4.88 & 4.88 &  $\underline{3.44}$ &  5.36 \\ 
			0.6 & 4.96 & 4.96 & 4.96 & 4.96 & $\underline{3.54}$ &  5.78 \\ 
			0.8  & $\underline{3.98}$ & $\underline{3.98}$ & $\underline{4.00}$ & $\underline{3.98}$ & $\underline{2.84}$ & 5.82 \\ 
			0.9 & $\underline{2.40}$ & $\underline{2.40}$ & $\underline{2.42}$ & $\underline{2.40}$ & $\underline{1.70}$ & 5.48 \\ 
			0.95 & $\underline{1.76}$ & $\underline{1.76}$ & $\underline{1.78}$ & $\underline{1.76}$ & $\underline{1.18}$   & 5.38 \\ \hline
			\multicolumn{7}{c}{{\textbf{Model {2}:} Polynomial decay}} \\	
			1.0 & 4.70 & 4.70 & 4.70 & 4.70 & $\underline{3.38}$  &  5.96 \\ 
			1.5  & 4.62 & 4.62 & 4.62 & 4.62 & $\underline{3.62}$ & 5.48 \\ 
			2.0  & 4.44 & 4.44 & 4.44 & 4.44 & $\underline{3.18}$ & 5.38 \\ 
			2.5  & 4.46 & 4.46 & 4.46 & 4.46 & $\underline{3.16}$  & 5.14 \\ \hline
			\multicolumn{7}{c}{\textbf{Model {3}:} Block-diagonal} \\	
			0.1  & 4.56 & 4.56 & 4.58 & 4.56 & $\underline{3.50}$ & 4.92 \\ 
			0.3  & 4.74 & 4.74 & 4.76 & 4.74  &   $\underline{3.74}$ & 5.32\\
			0.5 &  4.54 & 4.54 & 4.54 & 4.54 & $\underline{3.28}$ &  5.70 \\ 
			0.7  & $\underline{3.40}$ & $\underline{3.40}$ & $\underline{3.40}$ & $\underline{3.40}$ &  $\underline{2.46}$ & 5.62\\
			0.9 &  $\underline{1.88}$ & $\underline{1.88}$ & $\underline{1.92}$ & $\underline{1.88}$ &  $\underline{1.30}$ & 5.62\\ 
			\hline
			\hline
			
	\end{tabular}}
\end{adjustbox}	
\parbox{0.76\textwidth}{\footnotesize%
	\vspace{.1cm} 
	{Note}: We 	report the empirical FWERs (frequencies of falsely rejecting at least one hypothesis) of the controlling procedures. 
	The test statistics are generated from $N_d(\bm{0}, \bm{\Sigma})$ with different correlation structures. 
    The dimension $d$ is $100$ and the nominal significance level $\alpha$ is 5\%. The number of replications is $10^4$. 
	Instances with lower than 4\% FWER are underlined. 
}
\end{table}

\subsection{Under the alternative hypothesis}

Under the alternative hypothesis, the performance of the controlling procedures is assessed based on their global power (the percentage of replications that reject at least one hypothesis) and their successful detection rate (the percentage of overlapping hypotheses between the sets of false hypotheses  and discoveries). 
Given the improved accuracy of the SCC procedure in controlling the FWER under the null hypothesis (as demonstrated in Table \ref{tabSizeGlobalSequentialOtherMethods}), we anticipate that the SCC test will exhibit higher power when applied under the alternative hypothesis.
 
The test statistic vector $\bm{X}$ is generated from a $d$-variate normal distribution with mean vector $\bm{\mu} = (\mu_i)$  and a correlation matrix $\bm{\Sigma} = (\sigma_{ij})$, i.e., $N_d (\bm{\mu}, \bm{\Sigma})$. 
We adopt the same correlation matrices $\bm{\Sigma}$ as discussed in Section \ref{ssecAccuracy}.
The percentage of signals (i.e., non-zero $\mu_i$'s in the vector $\bm{\mu}$) is set to be 5\% (specifically, out of the 100 hypotheses, 5 are under the alternative).
All signals have the same strength $\abs{\mu_i} = \mu_0$ which is chosen to be relatively weak, i.e.,  $\pm2.1737$.\footnote{
	The chosen signal strength ensures that the test power  converges to unity as  $d \rightarrow \infty$ in the case of sparse signals, following the result presented in Theorem 3 of \citet[][]{liu2020cauchy}. 
	} 
The sign of the signal aligns with the sign of the test statistic under the null so that the signal always amplifies the magnitude of the test statistic.

\begin{table}[h] 
	\centering
	\caption{Global power (in\%) in a correlated setting with sparse signals}
	\label{tabPowerGlobalSequentialOtherMethods}
	\begin{adjustbox}{max width=\textwidth}
		\begin{tabular}{c cccccc}
			\hline
			\multicolumn{1}{c}{$\theta$}
			&\multicolumn{1}{c}{Bonferroni}
			&\multicolumn{1}{c}{Holm} 
			&\multicolumn{1}{c}{Hommel}  
			&\multicolumn{1}{c}{Hochberg} 
			&\multicolumn{1}{c}{Gumbel}
			&\multicolumn{1}{c}{SCC} \\
			\hline
			\multicolumn{7}{c}{{\textbf{Model 1:} Exponential decay}} \\
			0.2 & 66.50 & 66.50 & 66.62 & 66.50 & $\underline{59.32}$ & 76.20 \\ 
			0.4 & 66.48 & 66.48 & 66.60 & 66.48 & $\underline{59.86}$ & 76.84 \\ 
			0.6 & 65.70 & 65.70 & 65.74 & 65.70 & $\underline{58.58}$ &  75.82 \\ 
			0.8 & $\underline{63.86}$ & $\underline{63.86}$ & $\underline{63.90}$ & $\underline{63.86}$ & $\underline{56.90}$ & 72.74 \\ 
			0.9 & $\underline{60.62}$ & $\underline{60.62}$ & $\underline{60.78}$ & $\underline{60.64}$ & $\underline{53.90}$ & 68.42 \\ 
			0.95 & $\underline{57.24}$ & $\underline{57.24}$ & $\underline{57.30}$ & $\underline{57.24}$ & $\underline{50.58}$ & 63.94 \\ \hline
			\multicolumn{7}{c}{{\textbf{Model 2:} Polynomial decay}} \\
			1.0  & 66.04 & 66.04 & 66.14 & 66.04 & $\underline{58.76}$ &  74.82 \\ 
			1.5  & 65.88 & 65.88 & 66.00 & 65.88 & $\underline{58.46}$  &  75.10 \\ 
			2.0  & 65.88 & 65.88 & 66.00 & 65.88 & $\underline{58.84}$ & 75.20 \\ 
			2.5  & 65.22 & 65.22 & 65.28 & 65.22 & $\underline{58.22}$ & 74.66 \\ \hline
			\multicolumn{7}{c}{\textbf{Model {3}:} Block-diagonal} \\
			0.1 & 66.44 & 66.44 & 66.50 & 66.44 & $\underline{59.84}$ &   76.50 \\ 
			0.3  & 67.10 & 67.10 & 67.18 & 67.10 & $\underline{60.40}$  & 76.00 \\ 
			0.5  & 65.84 & 65.84 & 65.94 & 65.84 &$\underline{58.48}$ &  74.44 \\ 
			0.7 & $\underline{63.72}$ & $\underline{63.72}$ & $\underline{63.74}$ & $\underline{63.72}$ & $\underline{57.34}$ & 71.62 \\ 
			0.9 & $\underline{60.88}$ & $\underline{60.88}$ & $\underline{61.02}$ & $\underline{60.88}$ & $\underline{54.10}$ & 69.22 \\ 
			\hline
			\hline
	\end{tabular}}
\end{adjustbox}	
\parbox{0.76\textwidth}{\footnotesize%
	\vspace{.1cm} 
	{Note}: 
	We report the global powers (frequencies of rejecting at least one hypothesis), for various correlation matrices in the presence of sparse signals. 
	The test statistics are generated from $N_d(\bm{\mu}, \bm{\Sigma})$ with different correlation structures and sparse signals. 
	The dimension $d$ is fixed at $100$, and the percentage of signals is set to  $5\%$. All the signals have the same strength ($\pm2.1737$), with the sign depending on the sign of the test statistic under the null.
	We use a nominal significance level of 5\% and conduct $10^4$  replications. 
	We underline instances where the FWERs were lower than  4\% in Table \ref{tabSizeGlobalSequentialOtherMethods}. 
}
\end{table}

Table \ref{tabPowerGlobalSequentialOtherMethods}  shows the superior global power of the SCC test in the presence of correlated test statistics and sparse signals. 
The SCC test exhibits an approximate $10$\% power enhancement compared to the runner-up method. 
When the significance level $\alpha$ is set to $5\%$, the power of the SCC test ranges between  $69$\% and $77$\%. 
Although each statistical inequality-based method improves upon its predecessor in certain aspects, we do not observe a significant difference in the frequency of rejections among the four approaches. 
As anticipated, the Gumbel approach, which assumes independent test statistics,  is the most conservative test. 

Table \ref{tabPowerLocalSequentialOtherMethods} tells a similar story with respect to the average numbers of successful detections. The SCC testing procedure successfully detects approximately 1.2 out of 5 hypotheses (or $24\%$) under the alternative, even with sparse and weak signals.\footnote{The average number of false detections is slightly higher (not reported) for the SCC test, amounting to falsely rejecting on average 0.1 (out of 95) true hypotheses.} 
Meanwhile, the average number of successful detection for the  inequality-based procedures are around 0.95 (or $19\%$), whereas the value for the Gumbel procedure is about 0.81 (or $16.2\%$).

\begin{table}[!ht] 
	\centering
	\caption{Successful detection rates (in\%)  in a correlated setting with sparse signals}
	\label{tabPowerLocalSequentialOtherMethods}
	\begin{adjustbox}{max width=\textwidth}
	\begin{tabular}{c cccccc}
	\hline
	\multicolumn{1}{c}{$\theta$}
	&\multicolumn{1}{c}{Bonferroni}
	&\multicolumn{1}{c}{Holm} 
	&\multicolumn{1}{c}{Hommel}  
	&\multicolumn{1}{c}{Hochberg} 
	&\multicolumn{1}{c}{Gumbel}
	&\multicolumn{1}{c}{SCC} 
	\\
	\hline
	\multicolumn{7}{c}{\textbf{Model {1}:} Exponential decay}  \\
	0.2 & 18.80 & 19.00 & 19.00 & 19.00 & $\underline{16.00}$ & 23.20 \\ 
	0.4 & 19.20 & 19.20 & 19.20 & 19.20 & $\underline{16.20}$ & 23.60 \\ 
	0.6 & 18.80 & 19.00 & 19.00 & 19.00 & $\underline{16.00}$ & 23.40 \\ 
	0.8 & $\underline{19.20}$ & $\underline{19.40}$ & $\underline{19.40}$ & $\underline{19.40}$ &  $\underline{16.20}$ & 24.00 \\ 
	0.9 &  $\underline{19.00}$ & $\underline{19.00}$ & $\underline{19.20}$ & $\underline{19.00}$ & $\underline{16.00}$  & 24.00 \\ 
	0.95 &  $\underline{19.20}$ & $\underline{19.20}$ & $\underline{19.20}$ & $\underline{19.20}$ & $\underline{16.20}$ & 24.40 \\ \hline
	\multicolumn{7}{c}{\textbf{Model {2}:} Polynomial decay}  \\
	1.0 & 19.00 & 19.20 & 19.20 & 19.20 & $\underline{16.00}$  & 23.60 \\ 
	1.5 & 19.00 & 19.20 & 19.20 & 19.20 &$\underline{16.00}$  & 23.40 \\ 
	2.0 & 19.00 & 19.20 & 19.20 & 19.20 &$\underline{16.00}$ & 23.60 \\ 
	2.5 & 18.80 & 18.80 & 18.80 & 18.80 & $\underline{16.00}$ & 23.40 \\ \hline
	\multicolumn{7}{c}{
	\textbf{Model {3}:} Block-diagonal
	} \\
	0.1 & 19.40 & 19.40 & 19.40 & 19.40 & $\underline{16.40}$  & 23.60 \\ 
	0.3 & 19.40 & 19.60 & 19.60 & 19.60 & $\underline{16.60}$  & 23.60 \\ 
	0.5 & 19.40 & 19.40 & 19.40 & 19.40 &  $\underline{16.40}$ & 23.80 \\ 
	0.7 & $\underline{19.20}$ & $\underline{19.40}$ & $\underline{19.40}$ & $\underline{19.40}$ & $\underline{16.40}$  & 23.80 \\ 
	0.9 & $\underline{19.00}$ & $\underline{19.00}$ & $\underline{19.00}$ & $\underline{19.00}$ & $\underline{16.20}$  & 23.80 \\ 
	\hline
	\hline
	\end{tabular}}
	\end{adjustbox}	
	\parbox{0.76\textwidth}{\footnotesize%
	\vspace{.1cm} 
	{Note}: 
	We report the successful detection rates (overlap between the sets of alternative hypotheses and discoveries). 
	The data-generating processes used are the same as those for Table \ref{tabPowerGlobalSequentialOtherMethods}.
	We use a nominal significance level of 5\% and conduct $10^4$  replications. 
	We underline instances where the FWERs were lower than  4\% in Table \ref{tabSizeGlobalSequentialOtherMethods}.
}
\end{table}
 
\section{Example 1: Monitoring Drift Burst }
\label{secApplDriftBurst}

The drift burst hypothesis, as proposed by \citet{christensen2018drift},  postulates the existence of locally explosive trends in high-frequency asset prices, resembling phenomena like flash crashes or gradual jumps. 
An infamous example of a flash crash occurred on May 6, 2010, when the Dow Jones Industrial Average rapidly dropped nearly 1,000 points within minutes, only to recover most of the losses shortly thereafter. 

The drift burst test serves to detect and timestamp these explosive trends. 
In our analysis, we compute the test statistic on a minute-by-minute basis. Since there are 6.5 trading hours per day, the test needs to be conducted 341 times per day, resulting in a multiple testing problem. 
The test statistics are computed using overlapping windows and are expected to exhibit high autocorrelation. Given that drift bursts are rare events, they can be considered as sparse signals, and the strength of these signals can vary. 
There are several approaches to deal with the false discovery problem in this context, including the benchmark procedures considered in Section \ref{secSims}, the SCC, test and a resampling procedure suggested by \citet{christensen2018drift}.

Section \ref{ssecDB} presents  the drift burst hypothesis  and the drift burst test as a means to detect these phenomena. 
In Section 
S2.3
of the Online Supplement, we conduct a simulation study. We apply the controlling procedures to detect drift bursts in real-world data, specifically the Nasdaq composite index and S\&P 500 index ETFs, in Section \ref{ssecDBEmpirics}.
 
\subsection{Drift Burst Hypothesis and Test}\label{ssecDB}

 Under the null hypothesis of no drift burst, the frictionless log prices $P = (P_t)_{t \geq 0}$ follow an It\^o semi-martingale process defined on a filtered probability space $(\Omega, \mathcal{F}, (\mathcal{F})_{t \geq 0}, \Prob)$: 
\begin{eqnarray}
	\label{eqDGPNull}
	dP_t = \mu_t dt + \sigma_t dW_t,  
\end{eqnarray}
where $\mu_t$ is the instantaneous drift, and the diffusive component consists of the spot volatility $\sigma_t$ and a standard Brownian motion $W_t$.  The coefficients $\mu_t$ and $\sigma_t$ are locally bounded or ``non-explosive". The volatility process is assumed to follow a  \cite{heston1993closed}-type dynamics: 
\begin{equation*}
	d\sigma _{t}^{2}=\kappa \left( \omega -\sigma _{t}^{2}\right) dt+\xi \sigma
	_{t}dB_{t},
\end{equation*}
where $B_t$ is a standard Brownian motion and $E\left( dW_{t},dB_{t}\right) =r dt$.  

Under the alternative hypothesis, 
a drift-bursting term $\mu_t^\text{b}$ and a volatility-bursting component $\sigma_t^\text{b}$ are added to the standard It\^o semi-martingale process \eqref{eqDGPNull}, resulting in the following dynamics: 
\begin{eqnarray}
	\label{eqDGPAlt}
	d{P}_t =& 
	\mu_t dt 
	+ \sigma_t dW_t 
	+  \mu_t^\text{b }dt
	+ \sigma_t^\text{b} dW_t, 
\end{eqnarray}
where $\abs{\mu_t^\text{b}} / \sigma_t^\text{b} \rightarrow \infty$ as $t \rightarrow \tau_{\text{b}}$, with $\tau_{\text{b}}$ denoting the drift burst time.  An example of such an explosive process is given by: 
\begin{eqnarray}
	\label{simDB}
	\mu_t^\text{b} = 
	a \frac{\text{sign}(t-\tau_{\text{b}})}{\abs{\tau_{\text{b}}-t}^{\alpha^\text{b}}}
	\, \, \text{and} \, \, 
	\sigma_t^\text{b} = b \frac{\sqrt{\omega}}{\abs{\tau_{\text{b}} - t}^{\beta^\text{b}}},  
	\, \, \text{for} \, \, 
	t \in [\tau_\text{b} - \Delta t, \tau_\text{b} + \Delta t], 
\end{eqnarray}
where $2\Delta t$ represents the duration of the burst, $\alpha^\text{b}$ corresponds to the strength of the burst, $\beta^\text{b}$ is the strength of the volatility burst and $a$ and $b$ are constants. This data-generating process can capture various realistic patterns, including flash crashes and mildly explosive trends, and is used in the simulations in Section 
S2.3.

Observations are recorded at equidistant intervals $0 = t_0 < t_1 < \ldots <t_n = T$, where $T$ represents a fixed time period, such as one trading day consisting of 6.5 trading hours. The distance between two consecutive observations is $\Delta_n=t_{i+1}-t_{i}$. The observed log price, contaminated by noise, is denoted as  $\widetilde{P}_{t_i} = P_{t_i} + \epsilon_{t_i}$, where the  $\epsilon$ is an error term (noise) and independent from the latent log price $P$.

The noise-robust drift burst statistic  \citep{christensen2018drift} is defined as: 
\begin{eqnarray}
	\label{eqDBTestStat}
	X_i = 	\sqrt{h}	\frac{\hat{\bar{\mu}}_{t_i}}{\sqrt{\hat{\bar{\sigma}}_{t_i}^{2}}},
\end{eqnarray}
where $h$ is the bandwidth of the mean estimator, $\hat{\bar{\mu}}_{t_i}$ is a noise-robust estimator for the local drift, and $\hat{\bar{\sigma}}_{t_i}^{2}$ is a noise-robust estimator for the local variance. 
The test is applied on a coarse sampling grid, and the null hypothesis of the drift burst test  asserts that `there is no drift bursting at time period $t_i$'. 
The drift burst test statistics are computed minute-by-minute using overlapping rolling windows. As a result, the dimension of the test statistic sequence $\bm{X}=\left\{X_i\right\}$ is large (with the parameters settings we end up with $d=341$ for the period of one day) and the tests are autocorrelated. 
For more information regarding the implementation of noise-robust estimators of the local drift and local variance, we refer the reader to the Online Supplement. 

Under the null hypothesis, 
as the sampling frequency approaches infinity ($\Delta_n\rightarrow 0$), 
the test statistic \eqref{eqDBTestStat} converges to the standard normal distribution, i.e., $X_i \rightarrow^{d} N(0,1)$, 
This implies that the test statistic satisfies the necessary assumptions required by the Cauchy combination test when the sampling frequency is sufficiently high under the null. 
Under the alternative hypothesis, the test statistic diverges when the drift term explodes fast enough relative to the volatility, i.e., $\abs{X_i} \rightarrow \infty$, at the drift burst time. 

To control the familywise error rate,  \citet[][Appendix B]{christensen2018drift} propose a resampling-based approach for generating critical values for the drift burst test. 
The resampling  approach approximates the dependence structure of the test statistic sequence under the null hypothesis with an autoregressive order one (i.e., AR(1)) model and to obtain its distribution via simulation. 

There are a few limitations associated with using simulated critical values in practice.  Since each sequence of test statistics (corresponding to each day in an empirical analysis or each sample path in a Monte Carlo study) requires a unique critical value, the resampling procedure is computationally intensive.\footnote{To expedite the process, a table can be prepared in advance containing the quantile functions of the normalized maxima for various values of the autoregressive coefficient $\theta$ and dimensions $d$.  However, an interpolation routine becomes necessary when the estimated first-order autocorrelation and dimension are not included in the table.}  It also imposes a strong parametric assumption on the dependence structure, which could be misspecified. Additionally, resampling attempts to reproduce the dependence structure under the null hypothesis using possibly contaminated data \citep[as seen in the attenuation bias reported in][]{christensen2018drift},  
which in turn can affect the estimation of the critical value.
Therefore, caution should be exercised when interpreting the results, and the potential biases associated with the parameter estimation must be considered. 
See the Online Supplement for more detailed  discussion of the resampling approach.
  
In the Online Supplement, we also evaluate the performance of the aforementioned controlling procedures in a  simulation setting. Specifically, we evaluate the performance of the four inequality-based procedures, the Gumbel method, the resampling approach, and the SCC testing procedure. 
Instead of directly simulating the test statistics as done in Section \ref{secSims},  we generate log prices from \eqref{eqDGPNull} under the null hypothesis and \eqref{simDB} under the alternative hypothesis, and then compute drift burst test statistics \eqref{eqDBTestStat} from the simulated prices. 
Two types of drifts bursts are considered: a V-shaped 20-minute flash crash and a three-day persistent expansion. 

Our findings suggest that the SCC procedure is the most preferred option for monitoring drift bursts.  
The inequality-based methods and the Gumbel method tend to be conservative when applied to the autocorrelated drift burst test statistics. 
While the resampling procedure shows slightly better performance in terms of power and successful detection rate for flash crashes, the SCC procedure significantly outperforms the resampling method in identifying persistent expansions. 
Although there might be a small loss in power observed in some specific cases, it is a reasonable trade-off for having a robust approach to monitor drift bursts.
Moreover, the SCC procedure procedure offers the advantage of being easier to implement and can accommodate arbitrary dependency structures without requiring simulations, making it a more practical choice overall.

\subsection{Empirics}
\label{ssecDBEmpirics}

We apply  the same controlling procedures to the drift burst test 
using data from the Nasdaq ETF (ticker: IXIC) and  the S\&P 500 ETF (ticker: SPY) covering the period from 1996 to 2020.  
The data was obtained from the Refinitiv Tick History Database at a one-second frequency, and we follow the data cleaning rules outlined in \citet{barndorff2009realized}. 
We test for drift bursts on a minute-by-minute basis and control the familywise error rate at a level of $0.1$\%. 

\begin{figure}[!h] 
\caption{Drift bursts in the Nasdaq and S\&P 500 ETFs from 1996 to 2020}\label{figEmpiricalPrices}\centering	
	\subfloat[Nasdaq]{{\includegraphics[width=.38\textwidth,angle = -90]{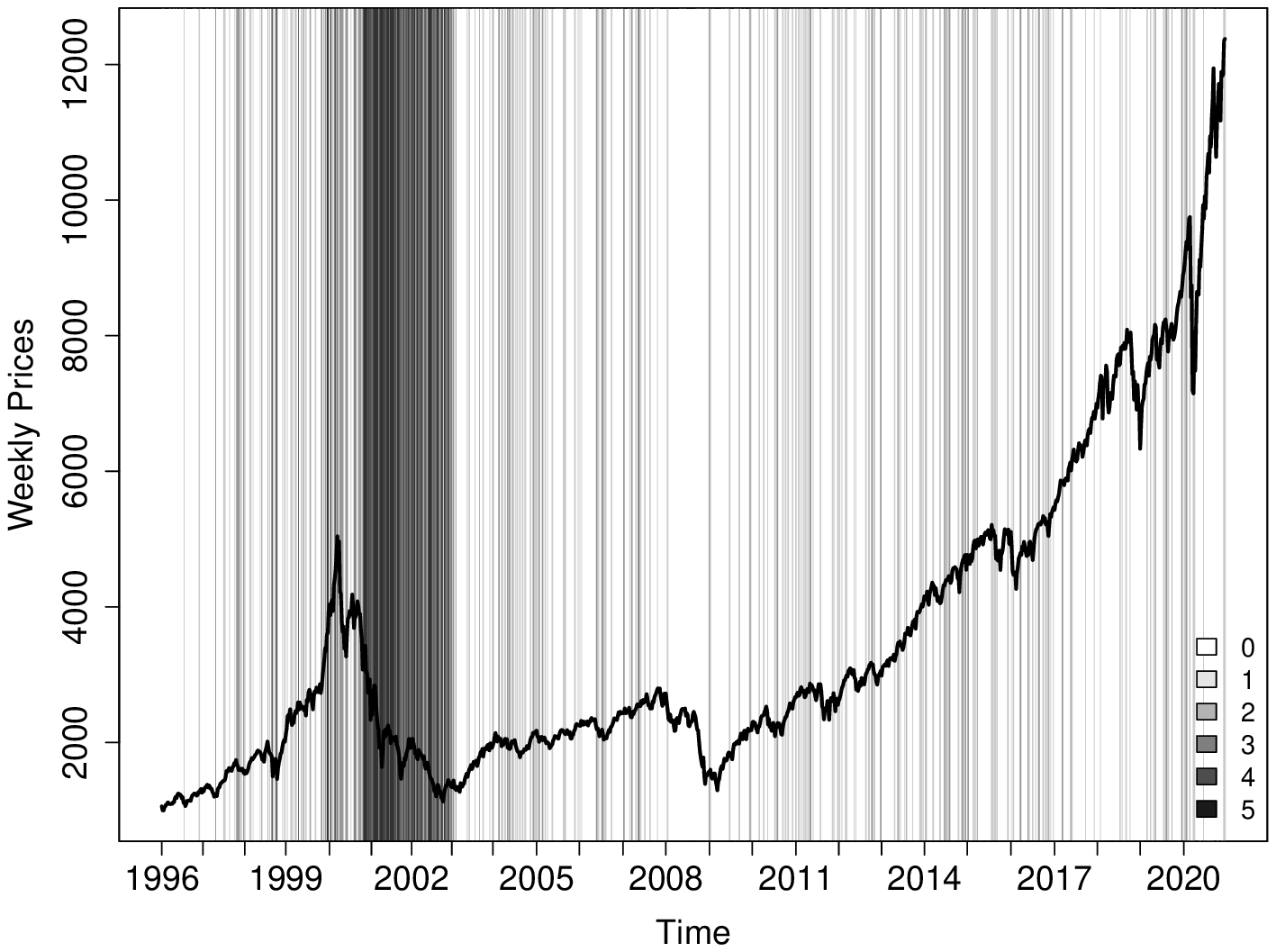} }}
	\subfloat[S\&P 500]{{\includegraphics[width=.38\textwidth,angle = -90]{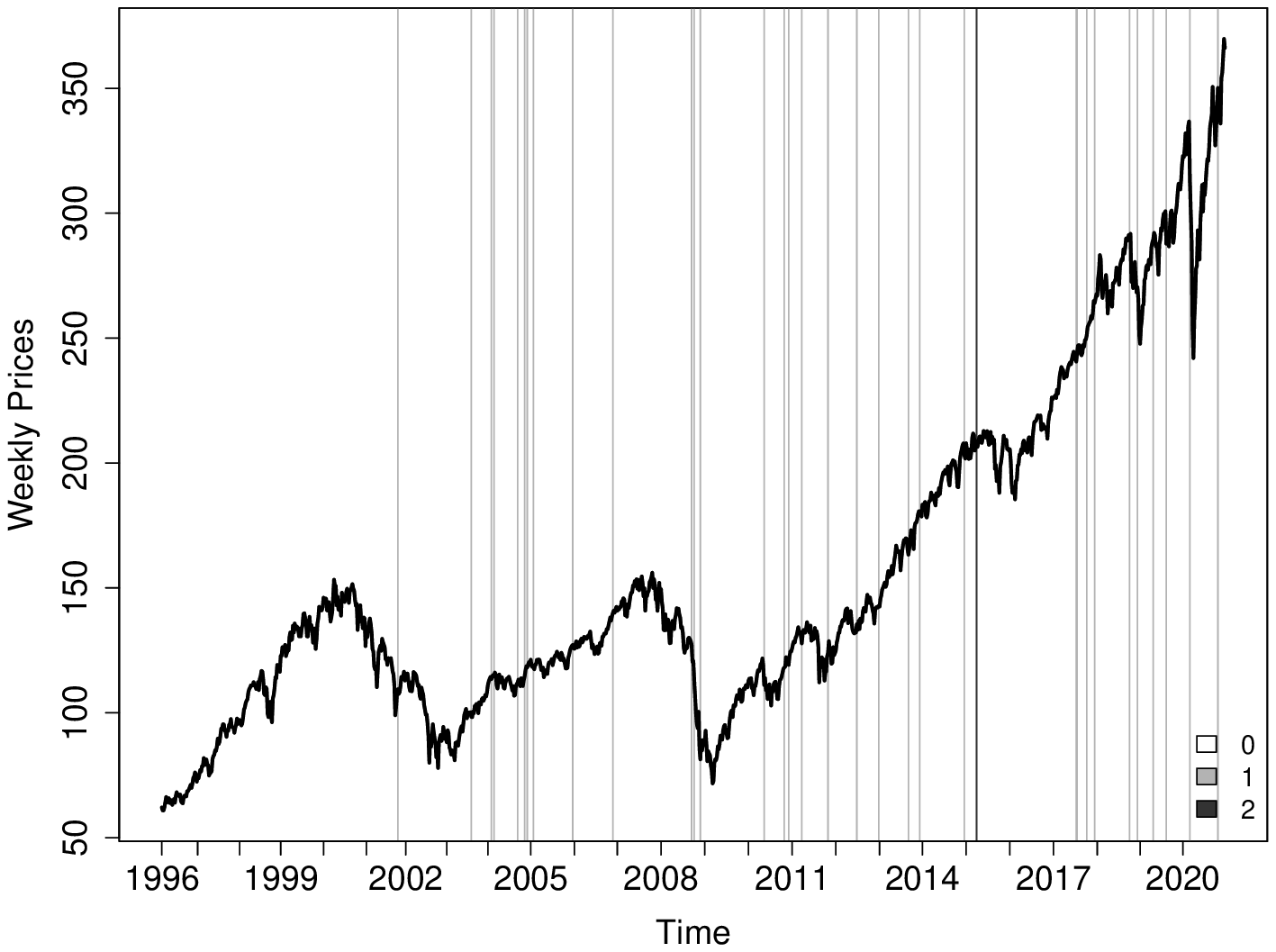} }}	\\
	\begin{minipage}{1.0\linewidth}
	\begin{tablenotes}
	\small
	\item {
		\medskip	
		Note: The black lines are the weekly index ETF prices. The grey bars indicate the number of rejections of the `no drift burst' null within the week.	The darker the grey scale the more drift burst days we observe. 
		We use the drift burst test of \cite{christensen2018drift} and the SCC testing procedure to control for the FWER at the $0.1\%$ significance level.
		}
	\end{tablenotes}
	\end{minipage}	
\end{figure}

The weekly prices of the Nasdaq and S\&P 500 ETFs are illustrated in Figure \ref{figEmpiricalPrices}, with grey bars indicating weeks that exhibit drift bursts. Drift bursts are determined using the test proposed by \cite{christensen2018drift} and the SCC controlling procedure. It is evident that the Nasdaq index experienced a higher number of drift bursts, particularly in the early 2000s following the collapse of the dot-com bubble. 
The S\&P 500 index barely has any rejections. 

Table \ref{tabEmpiricalR} reports the rejection frequencies of the drift burst test using the aforementioned controlling procedures. 
Compared to the benchmark procedures, the SCC test demonstrates a higher sensitivity in detecting drift burst days (i.e., at least one rejection within the day) and time intervals (i.e., the total number of rejections) in the Nasdaq index. 
In the case of the S\&P 500 index ETF, the SCC testing procedure detects more drift burst days and intervals than the inequality-based and Gumbel methods, albeit slightly fewer than the resampling procedure. 
It is noteworthy that the bursting episodes in the S\&P 500 are relatively short-lived, aligning with  the Flash Crash data-generating process considered in our simulations. On the other hand, the Nasdaq index exhibits  persistent drifts \citep[as reported in][]{laurent2022unit}, akin to the persistent expansion data-generating process. 

\begin{table}[H]
			\caption{Rejection frequencies of the drift burst test: Nasdaq and S\&P 500 index ETFs}\label{tabEmpiricalR}
	\scalebox{0.95}[0.95]{
	\begin{tabular}{lccccccc}
		\hline
		\multicolumn{1}{l}{} 
		&\multicolumn{1}{l}{Bonferroni} 
		&\multicolumn{1}{l}{Holm}
		&\multicolumn{1}{l}{Hommel}  
		&\multicolumn{1}{l}{Hochberg}  
		&\multicolumn{1}{l}{Gumbel} 
		&\multicolumn{1}{c}{Resampling} 
		&\multicolumn{1}{l}{SCC} 
		\\
		\hline
				\multicolumn{8}{c}{Nasdaq}  \\
		Days (in\%)   
		& 18.840
		& 18.840
		& 18.840 
		& 18.840
		& 11.662
		& 23.357
		& 23.522 
		\\
		Intervals (in\#) 
		&   6,479 
		&   6,540 
		&   6,559
		&   6,540
		&   3,823
		&   8,633
		& 10,944
		\\\hline
		\multicolumn{8}{c}{S\&P 500}  \\	
		Days (in\%)   
		& 0.526
		& 0.526
		& 0.526
		& 0.526
		& 0.247
		& 0.641
		& 0.576
		\\
		
		Intervals (in\#) 
		& 41
		& 41
		& 41
		& 41
		& 18
		& 55
		& 50
		\\	

		\hline
		\hline	
\end{tabular}}
\begin{minipage}{1\linewidth}
\begin{tablenotes}
\small
\item {
	\medskip
	Note: 
	The drift burst days are percentages of days over the full sample period with at least one rejection. 
	The figures in rows labelled intervals (in\#) are the total number of rejections over the entire sample period.
}
\end{tablenotes}
\end{minipage}
\end{table}

\section{Example 2: In Search of Nonzero Alpha Assets}
\label{secApplFan}

The Capital Asset Pricing Model (CAPM) is a prominent risk model. However, in light of empirical evidence revealing systematic patterns in stock
returns, often referred to as ``anomalies", many additional risk factors have been introduced to explain average returns. 
These risk factors are said to represent some
dimension of undiversifiable systematic risk that should be compensated
with higher returns. If the factor model fully characterizes expected
returns, the regression intercept (also known as the ``alpha") should theoretically equal zero. 
 
We search for
nonzero alpha assets using the \citet{fama2015five} five-factor model framework. 
The conventional approach is to run time series regressions on each individual asset, and subsequently perform individual tests on the estimated alphas. 
The number of assets that need to be tested simultaneously is large (e.g., all of the
S\&P 500 stocks) and the test statistics of the
cross-sectional alphas are most likely correlated due to the presence of unknown common factors 
\citep[see e.g.,][]{giglio2021thousands}. There is a general consensus in the empirical  finance literature that mispriced assets are rare \citep[see e.g.,][]{fan2015power,giglio2021thousands}. 
To tackle the challenge of multiple testing, several methods can be used. These include the benchmark procedures, the SCC test, and a screening procedure proposed by \citet{fan2015power}. All of these procedures control the FWER and have the ability to identify individual violations.\footnote{An alternative objective is to control the false discovery rate, which is defined as the proportion of false discoveries (see e.g., \citeauthor{barras2010false}, \citeyear{barras2010false} and \citeauthor{giglio2021thousands}, \citeyear{giglio2021thousands} for two examples).}

Section \ref{sec_HS} provides an overview of the nonzero alpha hypothesis and introduces the corresponding test.
Section 
S3.2 
of the Online Supplement presents simulation results that compare the performance of the controlling procedures in the context of the
nonzero alpha test. In Section \ref{sec_KF}, we
examine Fama-French portfolios formed on bivariate sorts and search for portfolios with a  nonzero alpha. 

\subsection{Nonzero Alpha Hypothesis  and Test}
\label{sec_HS}

The multi-factor pricing model, motivated by the Arbitrage Pricing Theory 
\citep{ross1976arbitrage}, postulates how financial returns are related to
market risks. This model has enjoyed widespread application in asset pricing and
portfolio management. 
Let $y_{it}$ be the excess return (i.e., real rate of return minus the
risk-free rate) of the $i$th financial asset at day $t$ and consider the
following linear regression model: 
\begin{eqnarray}  \label{eqCrossDecomp}
y_{it} = a_i + \bm{b}^{\prime }_i \bm{f}_t + u_{it}, \quad \text{with } i =
1, \ldots, d, \, t = 1, \ldots, T,
\end{eqnarray}
where $a_i$ is an intercept, often referred to as ``alpha", $\bm{b}_i = (b_{i1},
\ldots, b_{iK})^{\prime }$ is a vector of factor sensitivities or loadings, also known as ``betas", $\bm{f}_t = (f_{1t}, \ldots, f_{Kt})^{\prime }$ are
observable factors, and $u_{it}$ is an idiosyncratic error which is
uncorrelated with the factors. 
A well-known example of \eqref{eqCrossDecomp} is the three-factor model of \citet{fama1992cross}, 
which captures a substantial portion of the variation in the cross-section of average returns and absorbs a lot of the anomalies that have plagued the CAPM \citep[see
also][]{fama1996multifactor}.

Our objective is to identify individual assets with a nonzero alpha. The
null hypothesis of each asset is therefore $H_i: a_i=0$ (`there is no
mispricing of asset $i$') and the alternative hypothesis is $a_i\neq 0$
(`asset $i$ is mispriced'), for $i=1,\ldots, d$. 
The most common way to test
this null hypothesis is to use a simple $t$-statistic for $a_i$, i.e.,
\begin{eqnarray}  \label{eqTestAlpha}
{X}_i = \frac{\hat{a}_i}{ \hat{\sigma}_{\hat{a}_i}},
\end{eqnarray}
where $\hat{a}_i$ is the estimated alpha 
and 
$\hat{%
\sigma}_{\hat{a}_i}$ is the estimated standard error,  for each asset $i = 1, ..., d$. 
Under the null hypothesis, the test statistic \eqref{eqTestAlpha} follows a
Student-$t$ distribution, i.e., $X_i \sim t(\nu)$, with $\nu$ being the
degrees of freedom. 
A viable detection strategy involves computing the test
statistic \eqref{eqTestAlpha} for each asset in the cross-section and rejecting
the null hypothesis when $\abs{X_i}$ exceeds a pre-specified quantile of the 
$t(\nu)$ distribution.

The multiplicity issue arises when dealing with a large number of assets.  One important benchmark for controlling false discoveries in testing factor pricing models is the power enhancement test proposed by \cite{fan2015power}. 
This global test employs a screening technique that incidentally identifies individual violations. 
We refer to this approach as the screening method and provide a comprehensive description of the procedure in the Online Supplement. 
We can also use other approaches such as 
the  inequality-based, Gumbel method and SCC test.  
It is worth noting that the cross-sectional test statistics are likely to be
cross-correlated \citep[see e.g.,][]{giglio2021thousands}, and hence the Gumbel method and inequality-based procedures are again expected to be conservative. 
While it is true that a Student-$t$ distribution does not exactly fulfill the assumptions of the Cauchy combination test,  \citet{liu2020cauchy}
 show through simulations that the Cauchy approximation
remains accurate under such a departure from normality. 

Once again, we assess the finite sample performance of the SCC testing
procedure by comparing it with existing controlling procedures, which include the four inequality-based procedures, the Gumbel method and the screening approach, 
for the identification of nonzero alpha assets in both simulation settings and empirical applications. 
We exclude the resampling approach considered for the drift burst test, as it is not suitable for the current cross-sectional context.
Complete details of the simulation designs and results can be found in the Online Supplement.  
Unlike the simulations in Section \ref{secSims}, we simulate excess returns from the \citet{fama2015five} five-factor model, with its  parameters calibrated to the empirical data. We then compute the test statistics from the simulated data.
Our findings show that the SCC test outperforms all other procedures in terms of controlling the FWER, global power, and successful detection rate. The Gumbel method and screening method tend to be most conservative in their outcomes. 

\subsection{Empirics: Kenneth French Portfolios} \label{sec_KF}

In the empirical analysis, we study the portfolios 
available in
Kenneth French's Data Library.\footnote{\url{https://mba.tuck.dartmouth.edu/pages/faculty/ken.french/data_library.html}} 
Specifically, we focus on the  $d = 100$  portfolios formed  bivariately ($10\times10$) based on size and book-to-market  (Size-BM), size and investment (Size-INV), and size and operating profitability (Size-OP). 
The left-hand side of  equation \eqref{eqCrossDecomp} comprises value-weighted portfolio excess returns and the right-hand side are the five factors in the \citet{fama2015five} five-factor model (i.e., market, value, size, profitability and investment). 
The sample period spans from July 1963 to November 2022 and consists of $713$ monthly observations. 
To conduct the nonzero alpha test, we use a rolling window approach with a window size of $T = 240$ observations.  
We treat the $100$ portfolios as one family and control the familywise error rate at the $5\%$ level.

Table \ref{tabCrossFFK} reports the average rejection frequencies of zero alpha null (across the rolling analysis) using various controlling procedures. As anticipated, there are few violations. For instance, in the case of the Size-BM portfolios, the SCC test shows an average rejection frequency of 2.60\%. Overall, the SCC test consistently yields higher rejection frequencies compared to the inequality-based methods, the screening procedure, and the Gumbel method, in that order. These empirical results align with prior research on mispricing, confirming the rarity of nonzero alpha assets  \citep[see \textit{e.g.},][]{fama1996multifactor,fan2015power,giglio2021thousands}. Nonetheless, the SCC testing procedure can detect more of these rare violations. 

\begin{table}[!ht] 
	\centering
	\caption{Average rejection frequencies (in\%) of zero alpha null across a rolling analysis}
	\label{tabCrossFFK}
	\begin{adjustbox}{width=0.85\textwidth}
		\begin{tabular}{c | ccccccccc}
			\hline
			&\multicolumn{1}{c}{Bonferroni}
			&\multicolumn{1}{c}{Holm} 
			&\multicolumn{1}{c}{Hommel}  
			&\multicolumn{1}{c}{Hochberg} 
			&\multicolumn{1}{c}{Gumbel}
			&\multicolumn{1}{c}{Screening}
			&\multicolumn{1}{c}{SCC} 
			\\
			\hline
			Size-BM  
			& 1.25 & 1.26 & 1.27 & 1.26 & 1.09 & 1.09 & 2.60 \\
			Size-OP 
			& 0.53 & 0.53 & 0.53 & 0.53 & 0.51 & 0.61 & 0.75
			\\
			Size-INV  & 1.19 & 1.19 & 1.19 & 1.19 & 1.13 & 1.15 & 1.40 
			\\
			\hline
			\hline
			
	\end{tabular}}
\end{adjustbox}	
\parbox{0.85\textwidth}{\footnotesize%
	\vspace{.1cm} 
	{Note}: 
	We test the zero alpha null hypotheses within the Fama-French five-factor model framework using various controlling procedures. The nominal level is 5\% and the rolling window size is $T = 240$. 
}
\end{table}

\begin{figure}[!ht] %
	\caption{Number of rejected Size-BM portfolios from 1963 to 2022} 
	\label{figFMRejections}\centering
	
	\includegraphics[width=.5\textwidth,angle = -90]{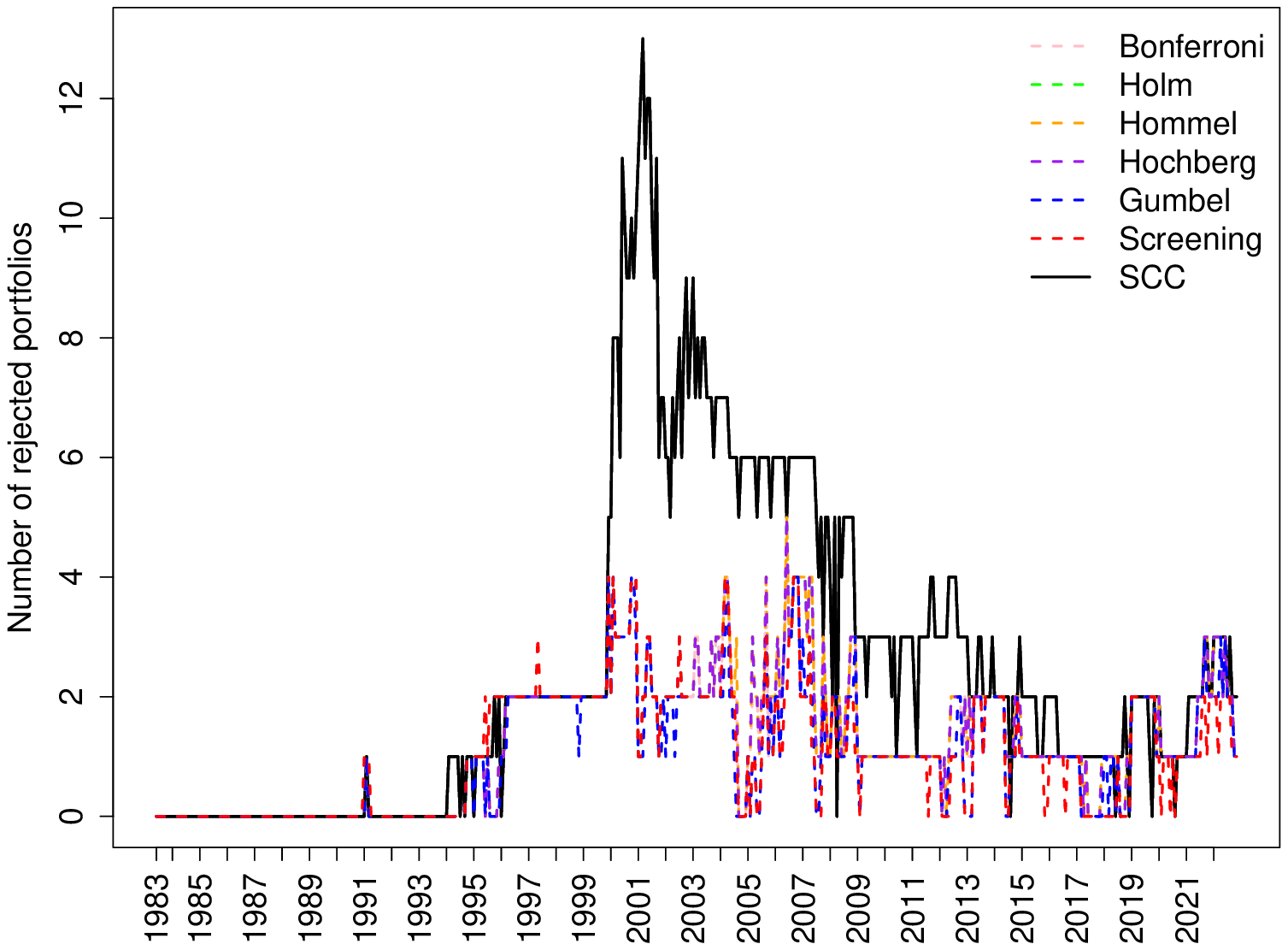}

	\begin{minipage}{0.8\linewidth}
		\begin{tablenotes}
			\small
			\item {
				\medskip
				Note: The nominal level is 5\% and the rolling window size is $T = 240$. 
			}
		\end{tablenotes}
	\end{minipage}
\end{figure}

Evidently, the rejection numbers vary over time,  as can be seen in Figure \ref{figFMRejections}. 
In the first decade of the sample period, there is almost zero rejection according to all procedures.  The number of identified nonzero portfolios starts to increase in the late 1990s, reaching its peak during the dot-com bubble crash in the early 2000s, and declined afterwards. 
The SCC test identifies more violations than other procedures for about $40\%$ of the sample period. 
During the remaining periods, the rejection numbers of the SCC test are on par with the other procedures. 
While the results from the benchmark procedures are similar, the gap between the SCC test and other procedures can be very substantial. For instance, in March 2001, the SCC test detected 13 portfolios with nonzero alphas, while the benchmark procedures identify only 1 or 2 such portfolios. 
The findings above align with our theoretical expectations and are consistent with our previous simulations results, reinforing the notion that using SCC can yield significant improvements in testing outcomes.

\section{Conclusions}
\label{secConc}

We introduce a simple procedure to control for false discoveries and identify individual signals in scenarios involving many tests, dependent test statistics, and  potentially sparse signals.
			The tool is agnostic to the underlying dependence structure and scalable to deal with high dimensions.
			 Our approach is a sequential version of the global Cauchy combination test proposed by  \cite{liu2020cauchy}. 
			 By applying the global test recursively on a sequence of expanding subsets of ordered $p$-values, 
			 our sequential Cauchy combination test enables the identification of individual violations.

We show that the sequential Cauchy combination test achieves strong familywise error rate control and is less conservative compared to  popular statistical inequality-based methods (such as the Bonferroni correction and  subsequent improvements of \citeauthor{holm1979simple}, \citeyear{holm1979simple},  \citeauthor{hommel1988stagewise}, 
\citeyear{hommel1988stagewise} and   \citeauthor{hochberg1988sharper}, 
\citeyear{hochberg1988sharper}) and the Gumbel method.

The Cauchy transformation has proven its value in a genome-wide association study of Crohn's disease \citep[][Section 4.3]{liu2020cauchy}, but its applicability extends beyond genomics. 
We revisit two important 
needle-in-a-haystack problems 
in financial econometrics, where the test statistics have either serial or
cross-sectional dependence:  monitoring drift bursts and searching for nonzero alpha assets. 
The drift burst test of \citet{christensen2018drift} detects the presence of explosive trends in asset prices using  high-frequency intraday data. The test statistics are computed from overlapping windows, resulting in high autocorrelation. 
We  also revisit the 
\citet{fama2015five}
multi-factor model to identify nonzero alpha financial assets. Detecting these rare nonzero alphas among a large group of financial assets is challenging, especially when the test statistics are likely to be cross-sectionally correlated.
Without a proper controlling procedure, one might flag false discoveries or miss important signals. Our results indicate that the sequential Cauchy combination test is the a preferable method for both applications.

We emphasize that our sequential Cauchy combination test is not limited to  financial econometrics. 
We anticipate its applicability to  a wide range of hypothesis tests in fields such as economics, finance, medicine, marketing and climate studies, as it can handle various types of dependence effectively.

\bibliographystyle{chicago}
\bibliography{Reference}

\appendix

\section{Proof of Theorem \ref{thm}}
\label{sec:proof}
\textbf{Proof.}	Let $\mathcal{R}^{(s)}$ be the collection of rejected hypotheses in step $s$
	with $s=\left\{ 1,2,\ldots ,d\right\} $, $\mathcal{R}^{(0)}=\emptyset$, and the rejection set $\mathcal{R}=%
	\mathcal{R}^{(d)}$. The SCC testing procedure can be viewed as a sequential rejection procedure, as outlined in Table \ref{tabDecisionRule}.
	\begin{table}[H]
		\caption{Sequential rejection procedure of the SCC test}
		\label{tabDecisionRule}
		\centering
		\begin{tabular}{p{1.cm}p{3.8cm}p{10.5cm}}
			\hline
			Step & Hypothesis & Decision \\ 
			$s=1$ & $\mathcal{H}_0^{\left( 1\right) }= \bigcap_{j=1}^d H_{(j)}$ & $%
			\mathcal{R}^{(1)}=H_{(1)}$ if $\widetilde{p}_{(1)}\leq\alpha$; otherwise $%
			\mathcal{R}^{(1)}=\emptyset $ \\ 
			$s=2$ & $\mathcal{H}_0^{\left( 2\right) }=\bigcap_{j=2}^d H_{(j)}$ & $%
			\mathcal{R}^{(2)}=\mathcal{R}^{(1)} \cup H_{(2)}$ if $\widetilde{p}%
			_{(2)}\leq\alpha$; otherwise $\mathcal{R}^{(2)}=\mathcal{R}^{(1)}$ \\ 
			$\ldots$ & $\ldots$ & $\ldots$ \\ 
			$s=d$ & $\mathcal{H}_0^{\left( d\right) }= H_{(d)}$ & $\mathcal{R}^{(d)}=%
			\mathcal{R}^{(d-1)} \cup H_{(d)}$ if $\widetilde{p}_{(d)}\leq\alpha$;
			otherwise $\mathcal{R}^{(d)}=\mathcal{R}^{(d-1)}$ \\ \hline
		\end{tabular}
	\end{table}
	Let $\mathcal{N}_{s}\left(\mathcal{R}^{(s-1)}\right)$ be the successor
	function, which represents hypotheses to be rejected at the step $s$ given $%
	\mathcal{R}^{(s-1)}$. The successor function of the SCC test is defined by: 
	\begin{equation*}
		\mathcal{N}_{s}\left( \mathcal{R}^{(s-1)}\right) =\left\{ 
		\begin{array}{ll}
			H_{(s)} & \text{if }\widetilde{p}_{(s)}\leq \alpha _{\mathcal{R}
				^{(s-1)}}=\alpha \\ 
			\emptyset & \text{otherwise}%
		\end{array}
		\right. ,
	\end{equation*}
	for $s=1,... ,d$, which is either an empty
	set or contains a single hypothesis. The rejection set at step $s$ is the
	union of the rejection set at the previous step $\mathcal{R}^{(s-1)} $ and
	its successor function $\mathcal{N}_{s}(\mathcal{R}^{(s-1)})$. In other
	words, we have  that the rejection set is made up of: 
	\begin{equation}
		\mathcal{R}^{(s)}= \mathcal{R}^{(s-1)} \cup \mathcal{N}_{s}(\mathcal{R}%
		^{(s-1)}) =\mathcal{N}_{1}(\mathcal{R}^{(0)}) \cup \mathcal{N}_2(\mathcal{R}%
		^{(1)}) \cup ...\cup \mathcal{N}_{s}(\mathcal{R}^{(s-1)}).
	\end{equation}
	This means that at each step of the procedure, we update the rejection set 
	by including the previously hypotheses and any new 
	hypotheses that would be rejected based on the successor function. By
	doing so, we accumulate evidence against the null hypotheses, and the
	rejection set either remains the same or grows as we progress.
	
	Suppose that $\mathcal{R}^{(s)}=\mathcal{F}$,  the probability of the SCC test making
	at least one false positive rejection is given by: 
	\begin{eqnarray*}
		\Pr \left\{ \mathcal{R}\cap \mathcal{T\neq \emptyset }\right\}  &=&\Pr
		\left\{ \mathcal{R}^{\left( d\right) }\cap \mathcal{T\neq \emptyset }%
		\right\}  \\
		&=&\Pr \left\{ \left( \mathcal{R}^{\left( s\right) }\cup \mathcal{N}%
		_{s+1}\left( \mathcal{R}^{\left( s\right) }\right) \ldots \cup \mathcal{N}%
		_{d}\left( \mathcal{R}^{\left( d-1\right) }\right) \right) \cap \mathcal{%
			T\neq \emptyset }\right\}  \\
		&=&\Pr \left\{ \left( \mathcal{N}_{s+1}\left( \mathcal{R}^{\left( s\right)
		}\right) \ldots \cup \mathcal{N}_{d}\left( \mathcal{R}^{\left( d-1\right)
		}\right) \right) \cap \mathcal{T\neq \emptyset }\right\}  \\
		&=&\Pr \left\{ \min_{j\in \left[ s+1,d\right] }\tilde{p}_{\left( j\right)
		}\leq \alpha \right\} =\Pr \left\{ \tilde{p}_{\left( s+1\right) }\leq \alpha
		\right\}. 
	\end{eqnarray*}%
	This is based on the definitions of the rejection set and the successor function and the facts that $\mathcal{F}\cap\mathcal{T}=\emptyset$ and $\tilde{p}%
	_{\left( j\right) }$ increases with $j$.  Since the null hypothesis
	corresponding to\ $\tilde{p}_{\left( s+1\right) }$ is $\mathcal{H}%
	_{0}^{(s+1)}=\bigcap_{j=s+1}^{d}H_{(j)}$, which equals $\mathcal{T}$ when $%
	\mathcal{R}^{(s)}=\mathcal{F}$, from \cite{liu2020cauchy}, 
	\begin{equation*}
		\lim_{\alpha \rightarrow 0}\Pr \left\{ \tilde{p}_{\left( s+1\right) }\leq
		\alpha \right\} \rightarrow \alpha 
	\end{equation*}%
	under Assumption \ref{ass1} when $d$ is fixed and Assumptions \ref{ass1} and \ref{ass2} when $d=o(h^{\eta})$ with $0<\eta<1/2$. Consequently,%
	\begin{equation*}
		\lim_{\alpha \rightarrow 0}\Pr \left\{ \mathcal{R}\cap \mathcal{T\neq
			\emptyset }\right\} =\lim_{\alpha \rightarrow 0}\Pr \left\{ \tilde{p}%
		_{\left( s+1\right) }\leq \alpha \right\} \rightarrow \alpha \text{.}
	\end{equation*}%

\clearpage

\end{document}


\def\spacingset#1{\renewcommand{\baselinestretch}%
		{#1}\small\normalsize} \spacingset{1}

\if1\blind
{
\title{\bf Online Supplement to ``Sequential Cauchy Combination Test for Multiple Testing Problems with Financial Applications''\thanks{
 		$^{\ddag}$ Nabil Bouamara, Louvain Institute of Data Analysis and Modeling in economics and statistics, Universit\'{e} catholique de Louvain;  	Email: nabil.bouamara@uclouvain.be.\\
 		$^{\ddag\ddag}$ S\'{e}bastien Laurent, Aix-Marseille University (Aix-Marseille School of Economics), CNRS \& EHESS, Aix-Marseille Graduate School of Management -- IAE; Email: sebastien.laurent@univ-amu.fr.\\ 
 		$^{\ddag\ddag\ddag}$ Shuping Shi, Department of Economics, Macquarie University; Email: shuping.shi@mq.edu.au. 
}
\vspace{0.1cm}
}

\author{Nabil Bouamara$^{\ddag}$, S\'{e}bastien Laurent$^{\ddag\ddag}$, Shuping Shi$%
^{\ddag\ddag\ddag}$ \\}

\maketitle
} \fi

\if0\blind
{
	\bigskip
	\bigskip
	\bigskip
	\begin{center}
		{\LARGE\bf Online Supplement to ``Sequential Cauchy Combination Test for Multiple Testing Problems with Financial Applications''}
	\end{center}
	\vspace{2cm}
} \fi


		\begin{abstract}
		This online appendix consists of three sections. Section \ref{ssecBenchmarks} gives an overview of the benchmark procedures, which includes four inequality-based methods and two procedures based on the maximum of the test statistics. 
		Section \ref{AppDB} complements Section 
		4 
		in the main text by providing additional information on the noise-robust estimators used in the drift burst test, the implementation of the resampling procedure, and a simulation study comparing different multiple testing corrections when monitoring drift bursts. 
		Section \ref{AppAlpha} complements Section 
		5 
		in the main text by providing additional details on the screening procedure and includes a simulation study focused on the nonzero alpha test.
					
		\end{abstract}
	
\singlespacing 

\section{Benchmarks}
\label{ssecBenchmarks}

The benchmark multiple testing corrections controlling the familywise error rate can be classified into two categories: those based on statistical inequalities and  those based on the maximum of the test statistics. We discuss their theoretical behavior when confronted with correlated test statistics. 

\subsection{Procedures based on statistical inequalities}
\label{ssecOrderdPvals}

We review four popular statistical inequality-based procedures:  the Bonferroni correction and the subsequent improvements by  \citet{holm1979simple}, \citet{hommel1988stagewise} and \citet{hochberg1988sharper}. 
The Bonferroni correction is the most commonly used method, where the null  hypothesis $H_{(i)}$ is rejected if the corresponding $p$-value is smaller than a stringent critical value  $c_i = \alpha/d$  for $i=1,\ldots,d$. The threshold is the same for all $p$-values. 

\citet{holm1979simple} and \citet{hochberg1988sharper} order the $p$-values in an ascending order, i.e.,  $p_{(1)}\leq p_{(2)}\leq \ldots \leq p_{(d)}$ with  $H_{(1)},H_{(2)},\ldots,H_{(d)}$ corresponding to their respective null hypotheses, and set the thresholds based on their ranks, $c_i=\alpha / (d - i + 1)$ for $i=1,2,\cdots,d$. However, they differ in their rejection procedures. 
\citet{holm1979simple}'s method ``steps up" from the smallest $p$-value to the largest $p$-value. It is a pessimistic approach that  scans forward and stops as soon as a $p$-value fails to clear its threshold. In other words,  at any stage $i$, $H_{(i)} $ is rejected only if $p_{(i)}$ $\leq c_i$ and all preceding hypotheses $H_{(j)}$ with $j< i$ have been rejected. 
On the other hand, \citet{hochberg1988sharper}'s method ``steps down" from the largest $p$-value to the smallest $p$-value, comparing $p_{(i)}$ against the critical value $c_i $. It is an optimistic approach that scans backward and stops as soon as a $p$-value succeeds in clearing its threshold, rejecting all remaining hypotheses. 
By design, Hochberg's method will reject as many hypotheses as Holm's method. 

\cite{hommel1988stagewise} proposes a more intricate procedure that applies  \citet{simes1986improved}' global test to subsets of $p$-values  $\left\{ p_{\left(k\right) }\right\} _{ k = i }^{d}$, instead of relying  only on a single $p$-value $p_{\left(i\right)}$, and 
thus borrows power across hypotheses. 
Hommel's method rejects the hypotheses $H_{(i)}$ if $p_{(i)} \leq c_i = \alpha / j$, where 
$$
j=\argmax_{i\in[1,n]} \left\{i: p_{(n-i+k)} >  k \alpha  / i  \text{ for all } k = 1, \ldots, i\right\}.
$$ 
If the maximum does not exist, then all hypotheses are rejected. The procedure uses the closure principle \citep{marcus1976closed} to extend \citet{simes1986improved}'s global test, and makes statements about individual hypotheses. 
It has been shown that Hommel's method has higher power than Hochberg's method \citep{hommel1989}. 

Bonferroni and \citet{holm1979simple}'s methods are based on the first-order Bonferroni inequality, which states that given any set of events, the probability of the union of a set of events is smaller than or equal to the sum of their individual probabilities. 
Under the null hypothesis, 
the probability that at least one hypothesis $H_{(i)}$ has a  $p$-value $p_{(i)} \leq \alpha / d$ 
is bounded by $\alpha$: 
\begin{align}  \label{eqIneqPval}
	\Pr\left( \min_{i} p_{(i)} \leq \frac{\alpha}{d} \right) 
	= \Pr\left(\bigcup_{i =
		1}^{d} \left\{ p_{(i)} \leq \frac{\alpha}{d} \right\} \right) 
	&\leq
	\sum^{d}_{i = 1} \Pr \left( p_{(i)} \leq \frac{\alpha}{d}\right) \leq d
	\frac{\alpha}{d} \leq \alpha.
\end{align}
The Bonferroni \eqref{eqIneqPval} inequality 
does not make any specific assumption about the dependence between the $p$-values, but it provides protection against the ``worst-case" scenario where all events are independent and the rejection regions are disjoint. 
The inequality becomes an equality when all test statistics are independent, and it is a strict inequality when the hypotheses are dependent. 
In other words, the Bonferroni correction is  conservative when the $p$-values are correlated.  

The methods proposed by \cite{hochberg1988sharper} and \cite{hommel1988stagewise} rely on  \cite{simes1986improved}'s inequality.
According to \citeauthor{simes1986improved}' inequality, if a set of hypotheses $H_{(1)}, ...,H_{(d)}$ are all true, the probability of the joint event is: 
\begin{equation}  \label{eqSimes}
	\Pr\left(\bigcap_{i=1}^{d} \left\{p_{(i) }> \frac{i\alpha}{d}\right\}\right) \geq 1-\alpha.
\end{equation}
\citeauthor{simes1986improved}' inequality was initially developed for independent uniform $p$-values but it is applicable to a wide large of multivariate distributions. 
However, simulations conducted by \citet{simes1986improved} show that the  \citet{simes1986improved} test can be overly conservative for highly correlated multivariate normal statistics, although it is generally less conservative than the classical Bonferroni correction.

\subsection{Procedures based on the maximum of test statistics}
\label{ssecMaxTest}

Another category of controlling procedures involves the use of the maximum of a group of test statistics, denoted as $X_m = \max_{i} \abs{X_{i}}$, where $i = 1, \ldots, d$. This maximum statistic is used to set a stringent critical value that can be applied to each individual hypothesis, ensuring control over the familywise error rate. 
Specifically, when the individual test statistics are independent and follow a standard normal distribution under the null hypothesis, a normalized version of $X_m$ converges to a Gumbel distribution. The distribution of the normalized maximum statistic is given by: 
\begin{eqnarray}
	\frac{X_m - C_d}{S_d} 	\sim	G,
\end{eqnarray}
where $G$ has a cumulative distribution function of 
$\Prob (G \leq x) = \exp (- \exp (-x))$. The constants $C_d$ and $S_d$ are defined as: 
\begin{eqnarray}
	C_d = 
	\sqrt{2 \log d} 
	-
	\dfrac{\log \pi 
		+ 
		\log(\log d)
	}
	{2\sqrt{2 \log d}}
	\, \,  \, \text{and} \, \,  \, 
	S_d = \dfrac{1}{\sqrt{2\log d}}.
\end{eqnarray}
To reject the $i$th null hypothesis $H_{i}$, we compare the absolute value of $X_i$ against the threshold $G^{-1} (1 - \alpha) S_d + C_d$, where $G^{-1} (1 - \alpha)$ represents the $(1-\alpha)$ quantile of the standard Gumbel distribution.

For instance, the Gumbel critical value has been used as a multiple testing correction when testing for  jumps  in high-frequency asset returns \citep[][]{lee2007jumps}.
However,  when the  sequence of test statistics exhibits strong correlation, the number of tests severely overstates the effective number of independent copies within a given sample. The Gumbel critical values tend to be overly conservative in such cases \citep[see e.g.,][]{christensen2018drift}. 

Resampling-based methods loosen the assumption of independence by incorporating the specific dependence structure in the  dataset under consideration, resulting in less conservative testing outcomes compared to the Gumbel method and the inequality-based procedures. 
Depending on the empirical problem of interest, the resampling can be carried out by bootstrap, permutation, simulation, or randomization \citep[see e.g.,][for comprehensive discussions on resampling methods and testing procedures]{white2000,romano2005exact,romano2005stepwise,lehmann2005testing}. 
An example of the resampling-based approach can be found in Section
 \ref{secResampling}. However, it is important to note that resampling methods can  be computationally intensive and require strong parametric assumptions to be imposed on the underlying dependence structure.

\section{Example 1: Monitoring Drift Burst}
\label{AppDB}

In Section \ref{AppDBEstim}, we provide details on the drift and variance estimators used to construct the drift burst test statistic discussed in the main text. 
Section \ref{secResampling} introduces the resampling-based method proposed by \citet[][Appendix B]{christensen2018drift}, designed to control the familywise error rate when monitoring drift bursts. Section \ref{ssecDBSim} outlines a simulation study that compares different multiple testing corrections for the drift burst test.

\subsection{Drift and Variance Estimators}
\label{AppDBEstim}

To reduce the impact of market microstructure noise, \citet{christensen2018drift} use pre-averaged returns. These pre-averaged returns as computed as: 
\begin{eqnarray}
	\label{eqPreavRets}
	\Delta_i^n \bar{P} 
	= \sum^{k_n - 1}_{j = 1} 
	g_j^n 
	\, 
	\Delta^n_{i + j} \widetilde{P},
\end{eqnarray}
where $\Delta_i^n \widetilde{P} = \widetilde{P}_{t_i} - \widetilde{P}_{t_{i-1}}$ is the discretely sampled noise-contaminated log return over $[t_{i-1}, t_i]$,  $k_n$ is the pre-averaging window, and $g_j^n = g(j/k_n)$ is a weight function. 
For the construction of the drift burst statistic, \citet{christensen2018drift} specify the weight function as $g(x) = \min(x, 1-x)$ and set the pre-averaged window $k_n$ to $3$.

The noise-robust estimator for the drift $\mu_t$ is computed from the pre-averaged returns as:
\begin{equation}
	\label{eqHatMu}
	\begin{split}
		\hat{\bar{\mu}}_t^n 
		&= 
		\frac{1}{h_n} 
		\sum^{n-k_n+2}_{i = 1} 
		K \left( \frac{t_{i-1} - t}{h_n} \right) 
		\Delta^n_{i-1} \bar{P} ,
	\end{split}
\end{equation}
where $t \in [0,T]$, $h_n$ is a bandwidth parameter, and $K(.)$ is a kernel function. 

The noise-robust variance estimator $\hat{\bar{\sigma}}_t^{2,n} $ 
is a heteroscedasticity and autocorrelation consistent (HAC)-type statistic, accounting for dependence in the pre-averaged returns. It is computed as: 
\begin{equation}
	\label{eqHatSigma}
	\begin{split}
		\hat{\bar{\sigma}}_t^{2,n} 
		&= 
		\frac{1}{h'_n} 
		\Bigg\{
		\sum^{n-k_n + 2}_{i = 1} 
		\left[
		K 
		\left(
		\frac{t_{i-1} -t}{h'_n} 
		\right)
		\Delta^n_{i-1} \bar{P}
		\right]
		^2
		%
		%
		\\
		&\quad +  
		2 \sum^{L_n}_{L = 1}
		w 
		\left( 
		\frac{L}{L_n}
		\right)		
		\sum^{n-k_n-L+2}_{i = 1}
		K 
		\left(
		\frac{t_{i-1} -t}{h'_n} 
		\right)
		K 
		\left(
		\frac{t_{i+L-1} -t}{h'_n} 
		\right)
		\Delta^n_{i-1} \bar{P}
		\Delta^n_{i-1+L} \bar{P}		
		\Bigg\}
		, 
	\end{split}
\end{equation}
where $h'_n$ is a bandwidth parameter, $\omega(.)$ is a kernel function, and $L_n$ is the maximum autocorrelation lag length. 
The lag length is determined as $L_n = Q^* + 2 (k_n - 1)$, with $Q^*$ computed from the raw returns $(\Delta \widetilde{P})_{i = 0}^n$ using automatic lag selection. 

To avoid look-ahead bias, \citet{christensen2018drift} suggest for $K(.)$ a left-side exponential kernel, $K(x) = \exp (-\abs{x})$ with $x \leq 0$. 
In our implementation, we use a ten-minute 
bandwidth for the mean estimator 
($h_n$ = 600) 
and a  50-minute 
bandwidth ($h'_n = 5h_n$) 
for the volatility estimator. 

The kernel function $w: \mathbb{R}_+ \rightarrow \mathbb{R}$ chosen satisfies the properties $w(0)=0$ and $w(x) \rightarrow 0$ as $x \rightarrow \infty$. We use a Parzen kernel for $w(.)$, defined as: 
\begin{eqnarray*}
	w(x)
	= \left\{
	\begin{array}{ll}
		1 - 6x^2 + 6 \abs{x}^3, 
		& \text{for} \, \, 0 \leq \abs{x} \leq 1/2, \\
		2(1-\abs{x}^3), 
		&  \text{for} \, \, 1/2 < \abs{x} \leq 1, \\
		0, 
		& \text{otherwise.} \\
	\end{array}
	\right.
\end{eqnarray*}

\subsection{Resampling}
\label{secResampling}

Resampling aims to approximate the dependence structure of the test statistic sequence under the null hypothesis and obtain its distribution through simulation. Specifically, the dependence structure is approximated by a stationary Gaussian autoregressive process of order one given by:   
\begin{equation} 
	\label{eqXAR1}
	X_{i}=\theta X_{i-1}+\epsilon _{i}, 
	\, 
	\text{for } i = 1, \ldots, d,
	\text{ with }\abs{\theta}<1\text{ and }%
	\epsilon _{i}\overset{i.i.d.}{\sim }N(0,1-\theta^{2})\text{.}
\end{equation}
Under this specification,  the autocovariance function of $X_i$ is $cov\left( X_{i},X_{j}\right) =\theta^{\left\vert i-j\right\vert }$, exhibiting exponential decay. The autoregressive coefficient $\theta$ in \eqref{eqXAR1} can be replaced with an estimate obtained from the empirical test statistic sequence $\left\{X_{i}\right\} _{i=1}^{d}$ using  conditional maximum likelihood.

The resampling procedure involves 
generating $R$ sequences of test statistics from \eqref{eqXAR1}. 
For each sequence, $r=1,\ldots,R$,  the maximum value  $X_{m}^{r} = \max_{i = 1, \ldots, d} \abs{X^r_i}$, 
is computed, resulting in a collection of simulated maxima $\left\{ X_{m}^{r}\right\} _{r=1}^{R}$. The critical value of the two-sided drift burst test is then determined as the $1-\alpha $ quantile of these simulated maxima. 

\begin{figure}[!ht]
	\centering
	\caption{Simulated critical values for the drift burst test}
	\includegraphics[width=.6\textwidth,angle = 0]{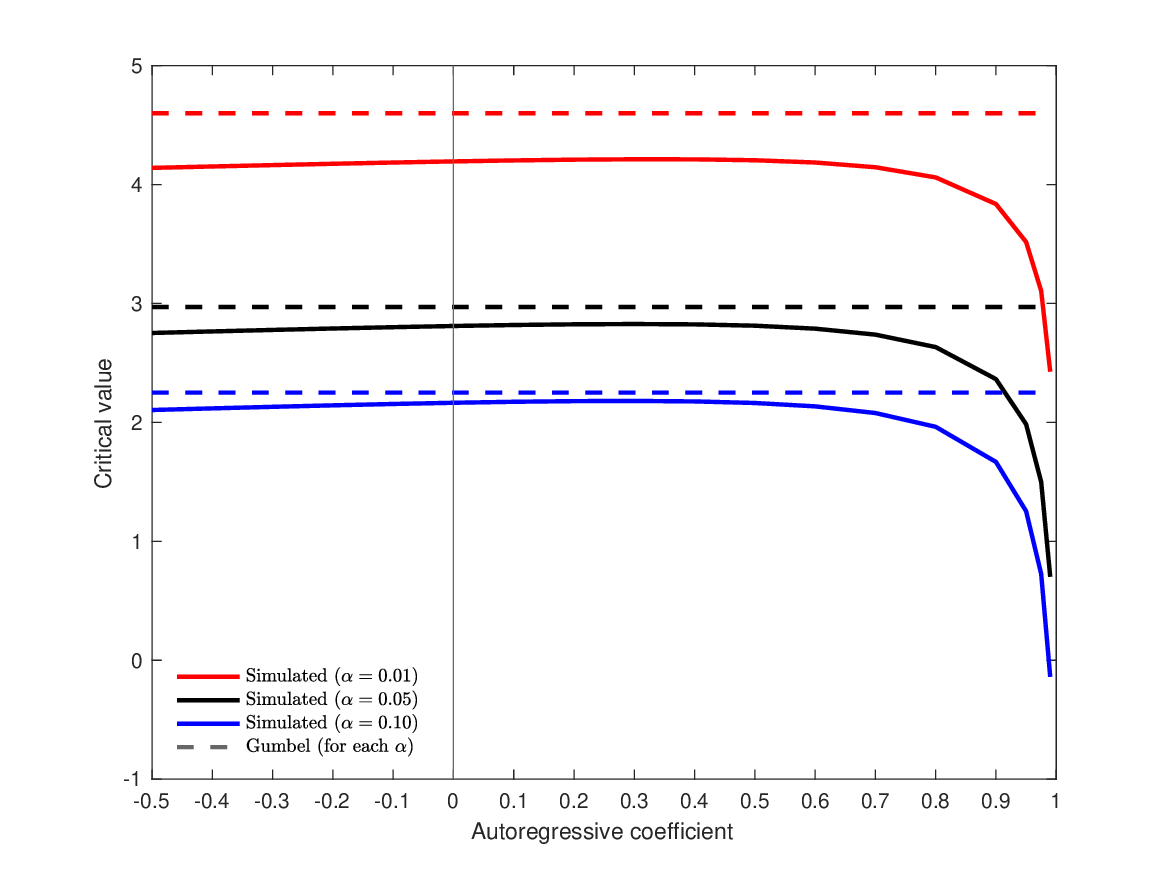}
	\label{figCOR18_fig8}%
	\begin{minipage}{1.0\linewidth}
		\begin{tablenotes}
			\small
			\item {
				\medskip
				Note: 
				This figure plots the (resampling-based critical values of the maxima when $d = 2,500$ and the Gumbel critical values. The figure shows how the varying degree of dependence (captured by $\theta$ ranging from $-0.5$ to $0.99$) affects the critical values at three different confidence levels. It extends \citet[][Figure 8]{christensen2018drift} to allow $\theta$ to take negative values.
				The simulation includes $R = 10^7$ replications with a burn-in of $10^4$ observations.
			}
		\end{tablenotes}
	\end{minipage}
\end{figure}

Figure \ref{figCOR18_fig8} plots the simulated critical values as a function of the autoregressive coefficient $\theta$ for $d = 2,500$ at three significance levels (1\%, 5\%, and 10\%), as in \citet[][Figure 8]{christensen2018drift}.
Comparing these simulated finite-sample critical values with those derived from the asymptotic Gumbel distribution, some observations can be made. 
When the autocorrelation is weak (i.e., $\theta \approx 0$) and the confidence level is set at $10\%$, the gap between the Gumbel and simulated critical values is relatively small. However, as we move towards extreme tail probabilities (e.g., $\alpha = 1\%$),  even when the test statistics are uncorrelated, the gap becomes more prominent.  
This behavior is expected since the convergence in law to the Gumbel distribution for the maximum term  is known to be slow \citep[as discussed  in][Appendix B, and references therein]{christensen2018drift}. 
The gap starts to widen noticeably in the region where the autoregressive coefficient exceeds $0.7$, a situation frequently encountered in rolling window implementations. 
The resampling-based method is expected to be less conservative than the Gumbel method (and the inequality-based methods) when the autocorrelation structure of $X_i$ follows an AR(1) process and when the estimate is accurate (i.e., $\hat{\theta}$ is close to $\theta$). This is because the resampling method takes into account the data-specific dependence structure. 

There are a few limitations associated with using simulated critical values in practice. 
The critical value obtained through resampling is unique for each sequence of test statistics. It is unique for each day in an empirical analysis and for each sample path in a Monte Carlo study. 
Obtaining these critical values is therefore more computationally demanding than other multiple testing corrections. 
To save time, a table can be prepared in advance, containing quantile functions of normalized maxima for different values of the autoregressive coefficient $\theta$ and dimensions $d$. 
However, when the estimated first-order autocorrelation and dimension are not included in the table, an interpolation routine becomes necessary. 

It is also important to note that the use of simulated critical value may be inappropriate to assess the strength of evidence against the null hypothesis.  As documented by \citet[][Appendix B, pp. 494]{christensen2018drift}, ``the estimated ACF [of an AR(1)]  is close to the observed one [for simulated data], although there is a slight attenuation bias for the empirical estimates". 
In practice, the estimation of the parameter $\theta$ in \eqref{eqXAR1} relies on empirical data that can be generated under either the null or alternative hypothesis.
This unavoidable mixture of test statistics under the null and the alternative hypotheses introduces  the possibility of biased autocorrelation estimates and simulated critical values, which can ultimately result in biased power calculations. 
Therefore, caution should be exercised when interpreting the results, and the potential biases associated with the parameter estimation must be considered.

\subsection{Simulation Study}\label{ssecDBSim}

We evaluate the performance of several multiple testing corrections for monitoring drift bursts in a  simulation setting. 
Specifically, we evaluate the performance of the four inequality-based procedures, the Gumbel method, the resampling approach, and the SCC testing procedure. 
Instead of directly simulating the test statistics as done in Section 
3, 
we generate log prices from 
(4.1)
under the null hypothesis and 
(4.3) 
under the alternative hypothesis, and then compute test statistics based on the simulated prices. 
Under the null hypothesis, we set $\mu_t^b=\sigma_t^b=0$ in Equation 
(4.2), 
while under the alternative we set $\mu_t^b$ and $\sigma_t^b$ as in 
(4.3). 

In line with \citet{christensen2018drift}, we use the following (annualized) parameters for the \citet{heston1993closed}-type variance process, namely $(\kappa, \omega, \xi, r) = (5, 0.0225, 0.4, -\sqrt{0.5})$. 
The parameter $\omega$ corresponds to a return volatility of roughly 20\% per annum. 
In each simulation, $\sigma_t^2$ is initiated at a random draw from its stationary law  $\sigma_t \sim \text{Gamma} (2 \kappa \theta \xi^{-2}, 2 \kappa \xi^{-2})$. 

Under the alternative hypothesis, we consider two different drift and volatility bursting patterns: 
\begin{enumerate}
	\item Flash Crash. 
	In this scenario, 
	the price experiences a brief, V-shaped flash crash lasting approximately $20$ minutes ($\Delta t = 0.025$) in the middle of the trading day ($\tau_{\text{b}} = 0.5$), following  \citet[][]{christensen2018drift}. Signals are sparse: the percentage of signals is about 6\% (20 out of 341 time intervals). 
	The strength of the signal is not constant, but  it reaches its peak at the bottom of the ``V" shape.  The drift burst rate is set to $\alpha^\text{b} = 0.65$ and the volatility burst rate is set to $\beta^\text{b} = 0.4$. The parameters $a$ and $b$ are set to $3$ and $0.15$, respectively. 
	The parameter choices generate a mild burst event.
	
	\item Persistent Expansion. 
	In this scenario, the price experiences  a three-day upward trend inspired by  \cite{laurent2020drift} and \cite{laurent2020volatility,laurent2022unit}. 
	The upward trend starts at the beginning of day one and culminates in a peak at the end of the third day. 
	The percentage of signals is 100\%, and the signal strength is progressively intensifies over the three days. 
	We set $\alpha^\text{b} = 0.75$, $\beta^\text{b} = 0.40$, $a = 3$, and $b = 0.15$. 
\end{enumerate}

The log prices are contaminated with a noise term $\epsilon_{t_i} \sim	N(0, n^{-1/2}\gamma\sigma_{t_i})$,  	where the  noise-to-volatility ratio $\gamma$ is fixed at $0.5$. 
The noise term is conditionally heteroskedastic and serially dependent (through $\sigma_{t_i}$). 

\begin{figure}[!h] 
	\centering
	\caption{A typical sample path of the DGPs}
	\label{figSamplePath}
	\subfloat[log Price with Flash Crash]{{\includegraphics[width=.5\textwidth,angle = 0]{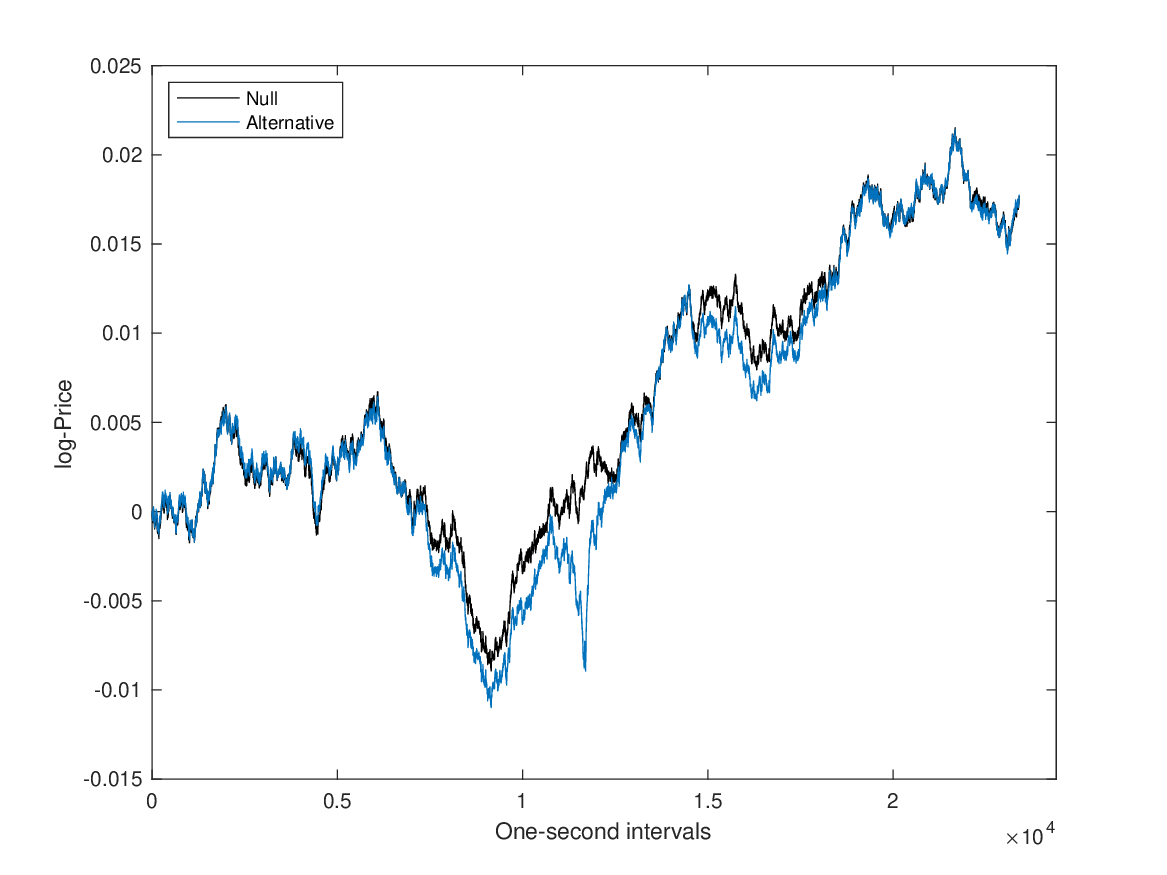} }}
	\subfloat[log Price with three-day rise]{{\includegraphics[width=.5\textwidth,angle = 0]{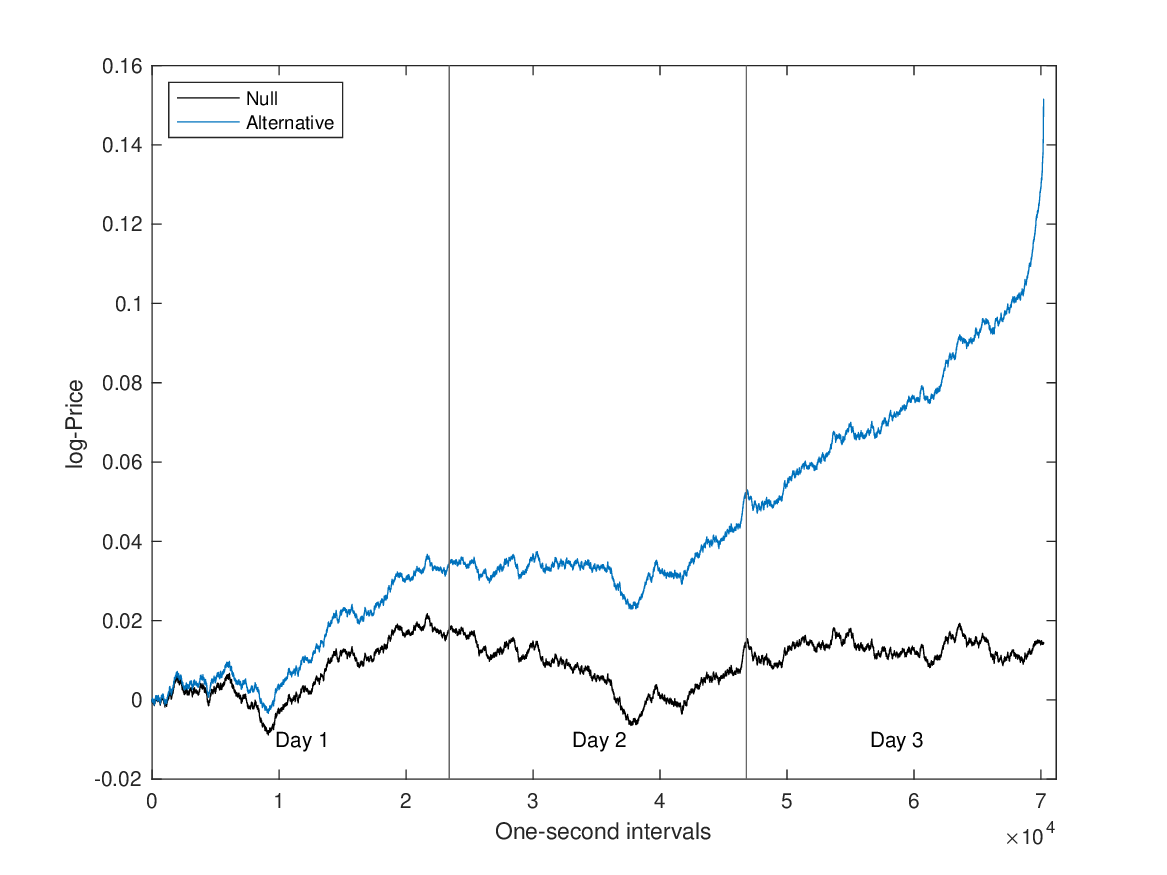} }}	\\
	\begin{minipage}{1.0\linewidth}
		\begin{tablenotes}
			\small
			\item {
				\medskip
				Note: 
				We plot a typical sample path of the DGPs 
				(4.1)
				 under the null and 
				 (4.2)
				 under the alternative with Flash Crash and Persistent Expansion patterns} 
		\end{tablenotes}
	\end{minipage}
\end{figure}

We simulate data at the one-second frequency, assuming 6.5 trading hours each day, resulting in a total of $23,401$ observations per day. 
Figure \ref{figSamplePath} plots a typical sample path of the data-generating processes.

We detect drift bursts on a one-minute grid. 
Each day is treated separately, with a burn-in period of 49 minutes, resulting in 341 tests per day. 
These 341 tests are treated as a family, and we  control the familywise error rate at 5\%. 
The simulation is repeated 2,000 times. 
Table \ref{tabSimsDriftBurst} reports the 
familywise error rate, global power, and successful detection rates
of the different controlling procedures, 
under the specified conditions. 
Additionally, the third column reports the average estimated first-order autocorrelation coefficient of the test statistics $\bar{\hat{\rho}}$. 
It is worth noting that there are moderate differences between the null and the alternative hypotheses, with a value of $0.894$ under the null and higher values under the alternative hypothesis. 
The difference becomes more pronounced  when the signal persists, such as $\bar{\hat{\rho}}=0.94 $ on the third day of the persistent expansion process.

\begin{table}[!ht] 
	\centering
	\caption{Finite sample performance of the controlling procedures for the drift burst test}
	\label{tabSimsDriftBurst}
	\begin{adjustbox}{max width=\textwidth}
		\begin{tabular}{l l | c | cccccccc}
			\hline
			\vspace{-.4cm} \\
			\multicolumn{2}{c}{DGP}
			&\multicolumn{1}{c}{$\bar{\hat{\rho}}$} 
			&\multicolumn{1}{c}{Bonferroni}
			&\multicolumn{1}{c}{Holm} 
			&\multicolumn{1}{c}{Hommel}  
			&\multicolumn{1}{c}{Hochberg} 
			& \multicolumn{1}{c}{Gumbel} 
			&\multicolumn{1}{c}{Resampling}
			&\multicolumn{1}{c}{SCC} \\
			\hline
			&& \multicolumn{8}{c}{Empirical FWER} \\ 
			\multicolumn{1}{l}{Heston}
			& 
			& 0.89
			& 2.40  &  2.40 &   2.40   & 2.40   & 1.80 &   4.80 &  5.25
			\\\hline
			&& \multicolumn{8}{c}{Global power} \\ 
			\multicolumn{1}{l}{Flash Crash} 
			&
			& 0.90
			& 67.70  & 67.70  & 67.70  & 67.70   & 65.45 &  74.45 & 69.75 
			\\
			\multicolumn{1}{l}{Persistent Expansion}
			& \text{Day 1}
			& 0.89
			& 5.90
			& 5.90
			& 5.90
			& 5.90
			& 4.75
			& 10.55
			& 15.55
			\\
			& \text{Day 2}
			& 0.89
			& 12.95
			& 12.95
			& 12.95
			& 12.95
			& 10.55
			& 19.85
			& 29.40
			\\
			& \text{Day 3}
			& 0.94
			& 100.00
			& 100.00
			& 100.00
			& 100.00
			& 100.00
			& 100.00
			& 100.00
			\\
			\hline
			&& \multicolumn{8}{c}{Successful detection rates} \\ 
			\multicolumn{1}{l}{Flash Crash} 
			& 
			& 0.90
			& 6.40  &  6.41  &  6.41  &  6.41 &   6.08  &  7.58 & 6.86
			\\ 
			\multicolumn{1}{l}{Persistent Expansion} 
			& \text{Day 1}
			& 0.89
			& 0.05
			& 0.05
			& 0.05
			& 0.05
			& 0.04
			& 0.11
			& 0.41
			\\
			& \text{Day 2}
			& 0.89
			& 0.13
			& 0.14
			& 0.14 
			& 0.14
			& 0.11
			& 0.25
			& 1.36
			\\
			& \text{Day 3}
			& 0.94
			& 11.20
			& 11.61
			& 12.06
			& 11.61
			& 10.73
			& 13.71
			& 23.59
			\\
			\hline
			\hline
	\end{tabular}}
\end{adjustbox}	
\parbox{\textwidth}{\footnotesize%
	\vspace{.1cm} 
	{Note}: 
	We consider different types of signals (Flash Crash and Persistent Expansion) and set the nominal level $\alpha$ to 5\%. The simulation is repeated $2,000$ times.
}
\end{table}

The SCC test and the resampling approach perform well under the null hypothesis, achieving  FWERs of $4.8\%$ and $5.25$\% respectively, at a nominal  level of 5\%. 
In contrast, the inequality-based  approaches and the Gumbel method are more conservative, with  empirical FWERs ranging between $1.8\%$ and $2.4$\% at a nominal level of 5\%.

Under the alternative hypothesis, the SCC test consistently outperforms the inequality-based approaches and the Gumbel method in terms of power and successful detection rate. 
When the signals persists through time, such as in the case of the Persistent Expansion, the SCC test exhibits higher global and local power  compared to the resampling approach. 
For instance, on the third day of the Persistent
 Expansion, the SCC test successfully detects approximately 24\% of intervals under the alternative, on average, while the resampling method only detects 14\% of intervals.
However, there are instances where the resampling approach shows higher power. 
When the duration of the signal is relatively short, such as in the case of the Flash Crash, the resampling approach performs better. 
Nonetheless, it is important to consider that a proper implementation of resampling is computationally demanding, and the critical values derived from it are susceptible to the attenuation bias (as discussed in \ref{secResampling}). 

Any potential contamination needs to be carefully considered when interpreting the results in the simulations (and  in the main text). 
Given the high autocorrelation of the drift burst test statistics, we are operating in the far-right region of Figure \ref{figCOR18_fig8}. 
Even minor downward or upward biased estimates of the autocorrelation under the null hypothesis can result in a loss or spurious power.  
Considering these challenges, we conclude that the SCC test is a more practical choice overall. It is easier to implement and can accommodate arbitrary dependency structures without requiring simulations. Although there might be a small loss in power observed in some specific cases, it is a reasonable trade-off for having a robust approach to monitor drift bursts. 

\section{Example 2: Nonzero Alpha Test}
\label{AppAlpha}

In Section \ref{ssecScreening}, we present the screening approach proposed by \cite{fan2015power}. 
In Section \ref{ssecSim}, we conduct a simulation study to assess the performance of the SCC test in  testing factor pricing models. 

\subsection{Screening}
\label{ssecScreening}

The power enhancement test proposed in \cite{fan2015power} is an important benchmark in testing factor pricing models. 
This global test examines whether there
is any asset with nonzero alpha, while also identifying individual violations as a byproduct. 
The global null hypothesis is that the alphas of the $d$ financial assets are jointly
indistinguishable from zero \citep[see e.g.,][]{fama1996multifactor}: 
$	\mathcal{H}_{0}:\bm{a} =\bm{0} \text{ with } \bm{a}=(a_{1},\ldots
,a_{d})^{\prime }.$

Classical tests for the global null hypothesis rely  on a quadratic statistic, which can present challenges in high-dimensional settings. 
Specifically, these tests typically have low power when faced with sparse alternatives. The accumulation of estimation errors under the null hypothesis causes the quadratic statistic to produce inflated critical values that overpower the signals present in the sparse alternatives \citep[see][for further
elaboration]{fan2015power}. 

The idea behind \citet{fan2015power}'s power
enhancement is to use a composite test statistic given by 
$J=J_{1}+J_{0}$, which combines a classical test statistic 
$J_{1}$ (with correct asymptotic size but low power under sparse alternatives) and a power-enhancing component $J_{0}$ designed to improve the power with minimal size distortion.  
An example of a power enhancement component is a screening statistic: 
 \begin{equation}
 	\label{eqScreeningStat}
	J_{0}=\sqrt{d}\sum_{j\in \widehat{S}}
	\frac{\widehat{a}_{j}^{2}}{\widehat{\nu }%
		_{j}},
\end{equation}%
in which $\widehat{\nu }_{j}$ is  the estimated asymptotic variance of $%
\widehat{a}_{j}$ and $\widehat{S}$ is a screening set containing the indices where the null hypothesis is violated. 

The power enhancing screening statistic \eqref{eqScreeningStat} is not intended to function as a stand-alone test. 
However, the screening set $\widehat{S}$ within  $J_0$ is the only part of the composite test $J$ that can incidentally identify individual violations \citep[as demonstrated in the empirical application of][Section 6.3]{fan2015power} and will therefore be included in our simulations and empirical application.

The screening set $\widehat{S}$ is defined as follows: 
\begin{equation*}
	\widehat{S}=\left\{ j:\frac{\abs{\widehat{a}_j}}{\widehat{\nu }_{j}^{1/2}}%
	>\delta _{d,T},j=1,\ldots ,d\right\}  \text{ with } \delta _{d,T}=C\log (\log (T))\sqrt{\log (d)}. 
\end{equation*}%
where $\delta_{d,T}$ is a threshold  chosen to dominate the maximum noise
level. In our implementation, we set the constant $C$ to $1.06$ as specified in the code shared by \citet{fan2015power} for replicating their simulations. 

The screening set is designed such that it  imposes a 0\% theoretical FWER, as the screening set is asymptotically empty under the null hypothesis. 
Under the alternative hypothesis, the screening test, which is based on the screening set, is expected to have power and enhance the power of a classical test statistic, specifically in the region 
$\left\{\max_{j\leq d}\frac{\abs{a_j}}{\nu_{j}^{1/2}}>3\delta _{d,T}\right\}$ (with $\nu_j$ being the asymptotic variance of $a_j$). 
For more details, we refer the reader to \cite{fan2015power}. 

\subsection{Simulation Study}
\label{ssecSim} 

We evaluate the performance of several multiple testing corrections for testing factor pricing models. 
Specifically, we evaluate the performance of the four inequality-based procedures, the Gumbel method, the screening approach, and the SCC testing procedure. 
Instead of directly simulating the test statistics as done in Section 
3, 
we generate excess returns from 
(5.1),
and then compute test statistics based on the simulated returns. 
We exclude the resampling approach discussed in  Section \ref{secResampling}, as it is not suitable for the current cross-sectional context. 

We generate excess returns $y_{it}$ from the 
five-factor model 
(5.1)
proposed by \citet{fama2015five}. We consider a setting with $d=100$ assets and $%
T=240$ time series observations, inspired by 
\citet[][Section 6.1]{fan2015power} and 
\citet[][Section 4.2]{shi2022relax}. 
Under the null hypothesis, all alphas are zero ($a_i=0$), for $i=1,\ldots,d$. Under the
alternative hypothesis, we introduce multiple  weak signals, where the alphas are defined as: 
\begin{eqnarray}\label{eq:ai}
	a_{i}=\left\{ 
	\begin{array}{cc}
		0.5 & \text{if }i\leq d^{0.4} \\ 
		0 & \text{if }i>d^{0.4}
	\end{array}
	,\right.
\end{eqnarray}
which matches the scenario considered in our empirical application in Section 
5.2.

The factor loadings and factor returns are generated from independent
multivariate distributions:  $%
\mathcal{N}_5 (\bm{\mu_B}, \bm{\Sigma_B})$ for the factor loadings and $\mathcal{N}_5 (\bm{\mu_f}, \bm{\Sigma_f})$ for the factor returns. 
We calibrate the parameters to the empirical Fama-French 100 portfolios that are sorted by Size-BM, Size-INV, and Size-OP, covering the period from January 1998 to December 2017. The choice of this sample period is guided by Figure 
4, 
where we observe substantial differences between the testing results during this timeframe. 
Table \ref{tabPars} reports the means and covariances for generating factor loadings $\bm{b}_i$ and factor returns $\bm{f}_t$. 
The mean vector ($\bm{\mu_f}$) and covariance matrix  ($\bm{\Sigma_f}$)  of the factor returns $\bm{f}_t$ are reported in the top panel. The mean vectors  ($\bm{\mu_B}$) and covariance matrices ($%
\bm{\Sigma_B}$) of  the factor loadings $\bm{b}_i$ for each of the different portfolio sorts are reported in the bottom panel. 

\begin{table}[H]
	\caption{Means and covariances for generating factor loadings $\bm{b}_i$ and factor returns $\bm{f}_t$}\label{tabPars}
	\centering
	\scalebox{1.}[1.]{
		\begin{tabular}{r | rrrrr }
			\hline
			\multicolumn{6}{c}{Parameters for factor returns} \\
			\hline
			\multicolumn{1}{c}{$\bm{\mu_f}$} 
			& \multicolumn{5}{c}{$\bm{\Sigma_f}$} 
			\\
			\hline
			$0.552$
			& $\phantom{-}19.942$ & $\phantom{-}3.472$ & $-1.742$ & $-6.725$ & $-3.185$
			\\ 
			$0.255$
			& $\phantom{-}3.472$ & $\phantom{-}10.183$ & $-0.613$ & $-4.774$ & $\phantom{-}0.198$ \\ 
			$0.145$
			& $-1.742$ & $-0.613$ & $\phantom{-}10.265$ & $\phantom{-}4.341$ & $\phantom{-}4.317$
			\\ 
			$0.315$
			& 	$-6.725$ & $-4.774$ & $\phantom{-}4.341$ & $\phantom{-}9.312$ & $\phantom{-}1.930$ 
			\\ 
			$0.245$
			& 	$-3.185$ & $\phantom{-}0.198$ & $\phantom{-}4.317$ & $\phantom{-}1.930$ & $\phantom{-}4.641$\\ 
			\hline
			\multicolumn{6}{c}{Parameters for factor loadings} \\
			\hline
			\multicolumn{1}{c}{$\bm{\mu_B$}} 
			& \multicolumn{5}{c}{$\bm{\Sigma_B}$} \\
			\hline
			\multicolumn{6}{c}{Size-BM} \\
			$1.029$ 
			& $\phantom{-}0.015$ & $-0.016$ & $\phantom{-}0.014$ & $\phantom{-}0.018$ & \phantom{-}%
			$0.000$ \\
			$0.578$ 
			&	$-0.016$ & $\phantom{-}0.178$ & $-0.028$ & $-0.067$ & $-0.028$ \\
			$0.201$ 
			&	$\phantom{-}0.014$ & $-0.028$ & $\phantom{-}0.131$ & $\phantom{-}0.056$ & $\phantom{-}0.011$ \\
			$0.047$ 
			& $\phantom{-}0.018$ & $-0.067$ & $\phantom{-}0.056$ & $\phantom{-}0.097$ & $\phantom{-}$%
			$0.024$  
			\\
			$0.050$
			& $\phantom{-}0.000$ & $-0.028$ & $\phantom{-}0.011$ & $\phantom{-}0.024$ & $\phantom{-}0.047$ \\
			\multicolumn{6}{c}{Size-INV} \\
			$\phantom{-}1.015$ & $\phantom{-}0.016$ & $-0.012$  &  $-0.005$  &  $\phantom{-}0.002$  &  $\phantom{-}0.004$ \\
			$\phantom{-}0.564$ &  $-0.012$  &  $\phantom{-}0.182$  &  $\phantom{-}0.012$ &  $-0.044$  & $-0.031$ \\ 
			$\phantom{-}0.109$ &  $-0.005$  &  $\phantom{-}0.012$  &  $\phantom{-}0.021$  &  $\phantom{-}0.021$ &   $\phantom{-}0.004$ \\ 
			$\phantom{-}0.034$ & $\phantom{-}0.002$  & $-0.044$  &  $\phantom{-}0.021$  &  $\phantom{-}0.076$  &  $\phantom{-}0.018$ \\
			$\phantom{-}0.110$ & $\phantom{-}0.004$ &  $-0.031$  &  $\phantom{-}0.004$  &  $\phantom{-}0.018$  &  $\phantom{-}0.136$ \\
			\multicolumn{6}{c}{Size-OP} \\ 
			$\phantom{-}1.028$ & $\phantom{-}0.015$  &  $-0.010$  & $-0.009$ &  $-0.004$ &  $-0.002$ \\ 
			$\phantom{-}0.538$ &  $-0.010$ &   $\phantom{-}0.179$ &   $\phantom{-}0.012$  &  $\phantom{-}0.035$ &   $-0.005$ 
			\\ 
			$\phantom{-}0.159$ &  $-0.009$ &   $\phantom{-}0.012$  &  $\phantom{-}0.045$ &   $\phantom{-}0.037$ &  $-0.013$ \\ 
			$\phantom{-}0.089$ & $-0.004$  &  $\phantom{-}0.035$  &  $\phantom{-}0.037$ &   $\phantom{-}0.188$  &  $\phantom{-}0.006$ \\
			$\phantom{-}0.022$ &  $-0.002$ &  $-0.005$ &  $-0.013$  &  $\phantom{-}0.006$   & $\phantom{-}0.034$ \\
			\hline
			\hline	
	\end{tabular}}
				%
\end{table}

The idiosyncratic noises are generated from a multivariate normal distribution, i.e., $%
\mathcal{N}_d (0, \bm{\Sigma_u})$. To generate cross-sectional dependence in the idiosyncratic noise that resembles the one 
observed in the empirical application, we compute the covariance matrix $\bm{\Sigma_u}$ as the sample covariance matrix of the residuals obtained from the OLS estimation of the five-factor model. These sample covariance matrices have dimensions of $100\times100$ and are available upon request.\footnote{We thank the authors of \citet{shi2022relax} for generously sharing their calibration code.}  

We treat the 100 portfolios within each subgroup (Size-BM, Size-INV, and Size-OP) as a family and control the familywise error rate at the  5\% level. 
The simulation is repeated 2,000 times.
Table \ref{tabCrossSize} reports the familywise error rate, global power, and successful detection rates of the different controlling procedures, under the specified conditions.  
Additionally, the second and third columns show the minimum and maximum off-diagonal elements of the empirical correlation matrix of the test statistics across all replications, indicating strong cross-sectional dependence. 
Under this scenario, all procedures have reasonably good control of the familywise error rate. 
The inequality-based,  Gumbel and screening methods show FWERs below the nominal level of 5\%, with the Gumbel and screening  methods being the most conservatives. 
Note that, in the presence of strong cross-sectional dependence, the screening test is found to have an empirical rejection frequency of approximately $3\%$, which is higher than what is expected asymptotically (i.e., 0\%).  
Interestingly, the SCC testing procedure consistently achieves FWERs very close to the nominal level of 5\% in all three cases, and  outperforms all other procedures in terms of global power and successful detection rate. 

\begin{table}[!ht] 
	\centering
	\caption{Finite sample performance of the controlling procedures for the nonzero alphas test}
	\label{tabCrossSize}
	\begin{adjustbox}{max width=\textwidth}
		\begin{tabular}{c | cc | cccc cccccc}
			\hline
			& \multicolumn{1}{c}{$\widehat{\rho} (\min)$}
			& \multicolumn{1}{c|}{$\widehat{\rho} (\max)$}
			&\multicolumn{1}{c}{Bonferroni}
			&\multicolumn{1}{c}{Holm} 
			&\multicolumn{1}{c}{Hommel}  
			&\multicolumn{1}{c}{Hochberg} 
			&\multicolumn{1}{c}{Gumbel} 
			&\multicolumn{1}{c}{Screening}
			&\multicolumn{1}{c}{SCC}	\\
			\hline
			&& \multicolumn{8}{c}{FWER}\\
			Size-BM 
			& --0.41  &   0.71
			& 4.00    &  4.00   &   4.00   &   4.00     & 3.70   &   
			3.00 & 
			5.10
			\\
			Size-INV 
			&  --0.44    &    0.60
			& 4.65   &   4.65    &  4.65   &   4.65     & 4.15      &
			3.40 &
			5.70
			\\ 
			Size-OP 
			&	--0.37      &   0.60
			& 4.30  &    4.30     & 4.30   &   4.30    &  3.80      &
			3.45 &
			5.20
			\\
			&&\multicolumn{8}{c}{Global power}\\
			Size-BM 
			&  --0.41      &      0.71
			& 54.15   &  54.20   &  54.25   &  54.20  &   51.55    & 
			49.15 &
			58.75
			\\
			Size-INV 
			& --0.43      &     0.60
			& 	72.10    & 72.10  &   72.10  &   72.10  &   70.10    
			& 67.80
			& 76.40
			\\
			Size-OP 
			& --0.36      &      0.59
			& 83.40   &  83.40  &   83.40  &   83.40 &     81.75   & 80.25 & 88.05
			\\
			&&\multicolumn{8}{c}{Successful detection rates}\\
			Size-BM  
			&  --0.41      &      0.71
			& 15.49 & 15.56 & 15.59 & 15.57 & 14.56 & 13.62 & 17.67
			\\
			Size-INV 
			& --0.43      &     0.60
			& 	27.80    & 27.92  &   27.99 & 27.92 & 26.45 & 24.73 & 30.88
			\\
			Size-OP 
			& --0.36      &      0.59
			& 41.45   &  41.63  &   41.65  &   41.63 &     39.62   & 37.88 & 45.72
			\\
			\hline
		\end{tabular}
	\end{adjustbox}	
	\parbox{1\textwidth}{\footnotesize%
		\vspace{.1cm} 
		{Note}: The data-generating process is 
		(5.1)
		with $a_{i}=0$ for all $i=1,...,100$ under the null hypothesis and $a_i$ specified in \eqref{eq:ai} under the alternative hypothesis. 
		The sample size is $T=240$ and the number of assets is $d=100$. 
		The nominal level $\alpha$ of the tests is 5\%. The simulation is repeated 2,000 times. }
\end{table}

\newpage

\bibliographystyle{chicago}
\bibliography{Reference}